\documentclass[aps,prl,showpacs,twocolumn]{revtex4-2}
\usepackage{graphicx}
\usepackage{subfig}
\usepackage{tabularx}
\usepackage{bm}
\usepackage{amsmath}
\usepackage{esvect}

\providecommand{\abs}[1]{\left\lvert#1\right\rvert}

\providecommand{\ket}[1]{\lvert #1 \rangle}

\providecommand{\be}{\begin{equation}}
\providecommand{\ee}{\end{equation}}
\providecommand{\ba}{\begin{eqnarray}}
\providecommand{\ea}{\end{eqnarray}}

\usepackage{mathbbol}

\usepackage{tikz}
\usepackage{amsfonts,amssymb}
\usepackage{dsfont}
\usepackage{physics}
\usepackage{tocvsec2}

\usepackage{graphicx}
\usepackage{subfig}
\usepackage{tabularx}
\usepackage{bm}
\usepackage{amsmath}
\usepackage{esvect}

\usepackage{siunitx}
\usepackage{hyperref}
\usepackage[capitalize]{cleveref}

\providecommand{\abs}[1]{\left\lvert#1\right\rvert}

\providecommand{\ket}[1]{\lvert #1 \rangle}

\providecommand{\be}{\begin{equation}}
\providecommand{\ee}{\end{equation}}
\providecommand{\ba}{\begin{eqnarray}}
\providecommand{\ea}{\end{eqnarray}}
\usepackage{amsfonts,amssymb}
\usepackage{dsfont}
\usepackage{physics}
\usepackage{tocvsec2}
 \DeclareUnicodeCharacter{202F}{\,}
\hypersetup{colorlinks=true,linkcolor=blue,citecolor = blue, urlcolor=black}
\newcommand{\beq}{\begin{equation}}
\newcommand{\eeq}{\end{equation}}

\usepackage{appendix}

\begin{document}

\title{Resources for bosonic metrology: quantum-enhanced precision from a superselection rule perspective}

\author{Astghik Saharyan$^{1}$}
\author{Eloi Descamps$^{1}$}
\author{ Arne Keller$^{1,2}$}
\author{Pérola Milman$^{1}$ }
\email{corresponding author: perola.milman@u-paris.fr}

\affiliation{$^{1}$Université Paris Cité, CNRS, Laboratoire Matériaux et Phénomènes Quantiques, 75013 Paris, France}
\affiliation{$^{2}$ Département de Physique, Université Paris-Saclay, 91405 Orsay Cedex, France}

\begin{abstract}

Bosonic systems—particularly in quantum optics and atomic physics—are leading platforms for achieving quantum-enhanced precision in parameter estimation. By exploiting properties such as mode and particle entanglement, it is possible to attain precisions that surpass the shot-noise limit with respect to key resources like probe number or energy. Yet the mechanisms by which these bosonic resources enable quantum enhancement remain unclear. Consequently, the design of optimal probes and evolutions often relies on case-by-case analyses, where continuous- and discrete-variable regimes are treated separately and their connection is still unclear. We develop a comprehensive framework for quantum metrology that unifies all known precision-enhancement mechanisms based on bosonic systems. Our approach employs a superselection-rule–compliant representation of the electromagnetic field that explicitly includes the phase reference, thereby enforcing total particle-number conservation and bridging the discrete and continuous limits of quantum optics and symmetric massive systems. Within this unified formalism—of which established results emerge as special cases—we identify the distinct roles of mode and particle entanglement for quantum-enhanced precision. The framework further provides general measurement-optimization strategies for arbitrary multimode entangled probe states and naturally incorporates noise and non-unitary dynamics, ensuring applicability to realistic experimental conditions.
\end{abstract}
\vskip2pc 
 
\maketitle

The precision achievable in parameter estimation with independent or classically correlated probes is fundamentally limited by shot noise, which scales inversely with the square root of the available resources, such as energy or probe number. Quantum strategies, however, can surpass this bound, achieving up to quadratic improvements in precision for the same resource budget \cite{MacconeScience, PhysRevLett.96.010401}. Quantum metrological protocols exploiting such advantages are already being implemented across a wide range of physical platforms, with applications in fundamental physics \cite{LIGO, PhysRevA.70.023801, Magnetometry} and even in biological systems \cite{Bio, Bio2}.

Bosonic systems of diverse types - including single photons \cite{PhysRevLett.114.170802, MacconeNature, PhysRevLett.131.030801,descamps2023timefrequency}, intense fields \cite{PhysRevLett.100.073601, dowlingQuantumOpticalMetrology2008, polinoPhotonicQuantumMetrology2020, duivenvoordenSinglemodeDisplacementSensor2017, Fadel_2025, hollandInterferometricDetectionOptical1993a, kwonQuantumMetrologicalPower2022, pinel_ultimate_2012}, atomic ensembles \cite{RevModPhys.90.035005, Sinatra}, and the vibrational modes of trapped ions \cite{PhysRevLett.121.160502} or of massive oscillators \cite{opto} - can be prepared in quantum states that have been theoretically shown, and experimentally demonstrated \cite{Giovannetti:2011chh}, to enable quantum-enhanced precision, {\it i.e.}, overcoming the shot-noise. However, depending on the system, the contribution to precision enhancement due to modes, in one hand, and to particle's statistics on the other hand, can differ significantly. For states with well defined particle number where bosons are distributed across orthogonal modes, mode {\it and} particle entanglement are identified as the key resources enabling quantum enhanced precision  \cite{MacconeNature, PhysRevLett.131.030801, benattiSubshotnoiseQuantumMetrology2010,daltonNewSpinSqueezing2014a, PhysRevA.94.033817, braunQuantumenhancedMeasurementsEntanglement2018, PhysRevA.94.033817, PhysRevLett.133.260402, PhysRevX.10.041012, RevModPhys.90.035005, PhysRevA.94.033817}. In contrast, in the continuous-variable (CV) regime, where states do not necessarily have a definite photon number, single-mode states can yield quantum enhanced precision \cite{dowlingQuantumOpticalMetrology2008, polinoPhotonicQuantumMetrology2020, duivenvoordenSinglemodeDisplacementSensor2017, Fadel_2025, hollandInterferometricDetectionOptical1993a, kwonQuantumMetrologicalPower2022, pinel_ultimate_2012, Safranek_2019, LuizExp, zurek_sub-planck_2001,LuizSub,Dalvit_2006, pinel_ultimate_2012, PhysRevLett.100.073601, Zhuang_2020, LIGO2} - as for instance quadrature squeezed states -, though employing ancillary modes can render the implementation of parameter estimation protocols more practical \cite{PhysRevA.33.4033}. Finally, a general form of squeezing can also lead to quantum enhanced precision for different physical systems \cite{Maccone2020squeezingmetrology}.

Despite significant advances, a comprehensive framework unifying the various regimes of bosonic resources in quantum metrology remains missing. By this, we mean a compact, system-independent description from which the metrologically relevant features of bosonic experimental platforms—whether operating in the photon-number-resolved or continuous-variable limits, and across single- or multimode regimes \cite{polinoPhotonicQuantumMetrology2020,Fadel_2025}—can be consistently identified, together with their respective contributions to precision scaling. Such a framework would, in particular, allow one to address central questions in bosonic quantum metrology: how do mode and particle entanglement each contribute to surpassing the shot-noise limit? Can their respective roles be independently identified?   

In this Letter, we address these questions by introducing a unified framework that consistently characterizes bosonic resources and provides a practical tool for designing optimal parameter-estimation strategies for arbitrary quantum states. Our approach reveals how quantum-enhanced precision emerges either from mode and particle properties. A central element is the explicit treatment of the phase reference as a physical resource—typically implicit in CV descriptions—which restores particle-number conservation and clarifies the origins of quantum precision enhancement. This formulation encompasses all regimes of bosonic quantum metrology—single or multimode, continuous or discrete—and yields a compact analytical expression that remains valid for noisy, multimode entangled probes.

We begin by recalling the basic principles of parameter estimation and its associated precision bounds. A metrological protocol starts by preparing a probe - a quantum state $\ket{\psi}$ - whose evolution depends on the parameter of interest, $\kappa$. The estimation precision depends on the choice of probe state, on the parameter-dependent evolution, and on the measurement strategy. Its lower bound depends on the Fisher information (FI) ${\cal F}_{\kappa}$, with $\delta \kappa \geq 1/\sqrt{\nu{\cal F}_{\kappa}}$, with ${\cal F}_{\kappa}= \sum_i \frac{1}{P_{\kappa}(x_i)}(\frac{\partial P_{\kappa}(x_i)}{\partial \kappa})^2$, where $P_{\kappa}(x_i)$ is the probability of obtaining an outcome $x_i$ that is used to estimate $\kappa$ for a given measurement strategy and $\nu$ the number of measurements. The FI is upper-bounded by the quantum Fisher information (QFI) ${\cal Q}$ for small parameter variations \cite{PhysRevLett.134.010804}, so $\delta \kappa \geq 1/\sqrt{\nu{\cal Q}}$, that is the optimization of the FI over all possible measurement strategies. For pure-state probes and unitary evolutions $\hat U=e^{i\hat H\kappa}$, the QFI takes the simple form ${\cal Q}=4\Delta^2 \hat H$, computed using the probe state $\ket{\psi}$. For non pure states and non-unitary evolutions \cite{PhysRevResearch.2.033389, LuizMetroSqueeze, 2010NaPho...4..357K, PhysRevLett.132.193603, PhysRevA.106.062442}, an upper bound for the QFI can be established through purification protocols \cite{LuizMetro} followed by computing the QFI for such pure states.

The quantum aspects of metrology are particularly appealing when one compares the scaling of precision with some resource, that is usually considered to be the average energy, or the average number of probes \cite{Jarzyna_2015}. While in classical protocols this scaling of precision is limited by the shot-noise (${\cal Q} \propto \overline n$, where $\overline n$ is the intensity of the field or the average number of particles), in quantum metrology precision scaling can display a quadratic enhancement $({\cal Q} \propto \overline n^2$), which is called the Heisenberg scaling. 



Numerous strategies involving bosonic systems surpass the shot-noise limit across very different dynamics associated with a variety of parameters \cite{DEMKOWICZDOBRZANSKI2015345, ZGoldberg_2021} . In atomic systems, probes such as spin-squeezed, anticoherent and Dicke states have been employed in rotation sensing, while in quantum optics entangled photons and single-mode nonclassical states enable phase and displacement estimation \cite{daltonNewSpinSqueezing2014a, PhysRevA.68.033821, PhysRevLett.60.385, PhysRevA.98.032113, PhysRev.93.99, SteveGear, benattiSubshotnoiseQuantumMetrology2010, Mitchell, Magnetometry, LuizSub, Zurek, Dalvit_2006}. This diversity of methods raises fundamental questions: What unifying principles underlie these seemingly distinct approaches? 

We begin addressing this question by recalling a key formal tool —the Schwinger representation \cite{Schwinger2001}—which maps pairs of bosonic modes onto angular-momentum operators. For each mode $i$, creation and annihilation operators satisfy $\hat a_i^{\dagger}\ket{n_i}_i=\sqrt{n_i+1}\ket{n_i+1}_i$ and $[\hat a_i,\hat a_j^{\dagger}]=\delta_{i,j}$. One can then define $\hat J_x^{(i,j)}=(\hat a_i^{\dagger}\hat a_j+\hat a_i\hat a_j^{\dagger})/2$, $\hat J_y^{(i,j)}=i(\hat a_i\hat a_j^{\dagger}-\hat a_i^{\dagger}\hat a_j)/2$ and $\hat J_z^{(i,j)}=(\hat a_i^{\dagger}\hat a_i-\hat a_j^{\dagger}\hat a_j)/2$. This formalism describes symmetric states of $N$ two-level systems, such as those in Bose–Einstein condensates, and underlies the quantum-optical description of the Mach–Zehnder interferometer (MZI): phase shifts between modes $i$ and $j$ correspond to evolutions generated by $\hat J_z^{(i,j)}$, while beam splitters implement rotations $e^{i\theta\hat J_y^{(i,j)}}$, with $\theta$ related to the transmissivity.

Beyond interferometry, the Schwinger representation gains a broader significance when formulated within a superselection-rule–compliant (SSRC) framework for the multimode electromagnetic field, whose central principles we now outline.

The standard description of single-mode states of the electromagnetic field—here referred to as the CV representation—expresses them as superpositions of Fock states, $\ket{\psi}_1=\sum_{n=0}^{\infty}c_n\ket{n}_1$, where subscript $1$ denotes an arbitrary mode and $\sum_{n=0}^{\infty} |c_n|^2=1$. This description implicitly assumes the existence of a phase reference: without it, states must either possess a well-defined total photon number or be described by a statistical mixture $\hat \rho$ diagonal in the photon-number basis, satisfying $[\hat \rho, \hat n_1]=0$. In other words, the pure state $\ket{\psi}_1$ formally violates the photon-number superselection rule. To avoid such implicit assumptions—which are known to affect precision analyses in quantum metrology~\cite{PhysRevA.85.011801}—, one can instead adopt a superselection-rule-compliant (SSRC) representation of the field, in which the phase reference is made explicit and treated as a physical resource. The SSRC formalism will be central to our work, since it naturally unifies the single- and multimode CV regimes with the multimode, fixed-particle-number regime, as detailed in~\cite{SM} and developed in~\cite{descamps2024superselectionrules, descamps2025} in the context of quantum computation. The simplest pure state of $N$ photons in the SSRC representation:
\be\label{Psi}
\ket{\Psi}=\sum_{n=0}^N c_n\ket{n}_1\ket{N-n}_2,
\ee
where subscripts $1$ and $2$ denote orthogonal modes that serve as mutual phase references. State~\eqref{Psi} possesses a well-defined total photon number, ensuring that its coherence is invariant under the free evolution generated by $\hat H = \hbar \omega (\hat n_1 + \hat n_2)$~\cite{Note}. Introducing $n_{\rm max} \le N/2$, the CV limit corresponds to the regime $\sum_{n=0}^N |c_n|^2 n \ll n_{\rm max}$, which allows one to formally take $n_{\rm max} \to \infty$ (see~\cite{RevModPhys.79.555, descamps2025}). In this limit, $\ket{\Psi}$ is faithfully represented by $\ket{\psi}_1$, {\it i.e.}, the phase reference (here represented by the mode $2$) can be omitted and implicit. The CV limit considered here is analogous to that discussed in~\cite{RevModPhys.90.035005, PhysRevA.68.033821, PhysRevA.65.033619}, where the Holstein--Primakoff (HP) transformation~\cite{PhysRev.58.1098} is employed. In this regime, the expectation values of operators coincide, for $N \gg 1$, with those obtained within the SSRC representation. However, the two approaches differ in both form and interpretation: the HP transformation does not account for superselection rules and directly treats single-mode CV states such as $\ket{\psi}_1$, while in the SSRC framework the single-mode CV description arises only as an approximation, valid when the system exhibits a strong particle-number imbalance between the chosen modes (modes $1$ and $2$ in Eq.~\eqref{Psi}).

States~\eqref{Psi} are formally equivalent to the total angular-momentum eigenstates of $N$ symmetric, indistinguishable two-level bosons, thereby establishing a direct correspondence between discrete and continuous-variable regimes 	and across different bosonic platforms.

We now illustrate how this unified description enables a consistent analysis of the different limits of quantum metrology. As an example, consider a two-mode rotation, $e^{i\hat J_{\vec{n}}^{(1,2)}\xi}=e^{i\hat J_z^{(1,2)} \phi}e^{i\hat J_y^{(1,2)}\theta}$ corresponding to a combination of a phase shift and a beam splitter. In metrological applications, one may seek to estimate any of the parameters $\phi$, $\theta$, or $\xi$ above; here we focus, for example, on the case of rotations around $\vec{y}$ axis ($\beta=\pi/2$, $\varphi=\pi/2$, defined in End Matter, Eq. \eqref{eqvariance}), corresponding to estimating $\theta$.

In the CV limit ($N \gg 1$), the dynamics are restricted to the tangent plane of the Bloch sphere~\cite{RevModPhys.90.035005} (see~\cite{SM}), {\it i.e.}, $\theta \ll 1$. A compact unified expression valid in both the CV and fixed-$N$ regimes is obtained by identifying $\theta = q/\sqrt{N}$ (see Eq. \eqref{x} in End Matter). The corresponding precision in estimating $q$—or equivalently, $\theta$—then reads:
\be\label{rotation}
\delta \theta= \delta\left (\frac{ q}{\sqrt{N}} \right ) \geq \frac{1}{\sqrt{4\nu \Delta^2 \hat J_y^{(1,2)}}}.
\ee

The second inequality reduces in the CV limit to $\delta q \ge 1/\sqrt{4\nu\Delta^2\hat p}$ (see End Matter, Eq.~\eqref{limiteps}), consistent with the fact that rotations around $\vec{x}$ and $\vec{y}$ correspond, in this limit, to translations in quadrature phase-space~\cite{descamps2025}. Consequently, Eq.~\eqref{rotation}—which can be generalized to arbitrary rotation axes $\vec n$ (End Matter, Eq.~\eqref{eqvariance})—holds for all states and regimes. Importantly, making the phase reference explicit reveals that particle entanglement, often hidden in the CV representation, is required to surpass the shot-noise limit. Indeed, within the SSRC framework, the only particle-separable states are spin-coherent states (Fock states $\ket{N}$ in a given mode), reducing to coherent states in the CV limit~\cite{SM, descamps2024superselectionrules, descamps2025}), for which one can always find $\vec n$ such that $\Delta^2\hat J_{\vec n}^{(1,2)} \propto N$ (or $\overline n$ in the CV limit). Notice as well that while the estimation of $\theta$ can reach the Heisenberg limit - scaling as $1/N$ - the CV limit constraint to small rotations angles yields a QFI that scales linearly with the mean photon number of mode~$1$, $\overline n$ (see End Matter, where rotations around the $z$ axis are also analyzed). The SSRC representation thus provides a unified framework for quantum metrology, explicitly treating the two modes as mutual phase references: for balanced photon numbers, the total photon number $N$ is the relevant resource, whereas in the CV limit— {\it i.e.}, strong photon-number imbalance between modes—the dynamics effectively reduce to a single-mode evolution, with $\overline n$ as the relevant resource.

We now extend our analysis to the multimode regime, showing how the SSRC representation elucidates the respective roles of mode entanglement and particle statistics in precision enhancement. As an illustrative step, we first examine a three-mode state—corresponding to a two-mode configuration in the CV limit—before generalizing to an arbitrary number of modes: 

\be \label{general}
\ket{\psi}=\sum_{\substack{n_1,n_2=0 \\ n_1+n_2 \leq N}} c_{n_1,n_2}\ket{n_1}_1\ket{n_2}_2\ket{N-n_1-n_2}_3,
\ee
where modes $1$, $2$ and $3$ are part of the probe's state (that reduces to modes $1$ and $2$ with $3$ serving as a phase reference in the CV limit). State \eqref{general} evolves according to $\ket{\psi(\theta)}=\hat {\cal U}\ket{\psi}$, where $\hat {\cal U}=e^{i\hat H\theta}$ encodes a physical transformation that depends on a parameter $\theta$ to be estimated and we suppose for example, to illustrate our method, that $\hat H$ has the form ($\hbar =1$):
\begin{eqnarray}\label{H}
&&\hat H =  \chi \sum_{i< j\leq 3}\frac{\chi_{i,j}}{\chi}\hat J_{\vec{z}}^{(i,j)} = \chi \hat n_{\zeta,\varphi}\\
&&=\chi(\sin{\zeta}\sin{\varphi}~\hat n_1+\sin{\zeta}\cos{\varphi}~\hat n_2 +\cos{\zeta}~\hat n_3) \nonumber,
\end{eqnarray}

with $\hat n_i=\hat a^{\dagger}_i\hat a_i$ and $\chi$ a normalization factor. The question we now address is how the QFI depends on the probe’s fundamental resources—its modal structure and particle statistics. To this end, we identify a collective operator $\hat n_{\zeta,\varphi}$ that maximizes ${\cal Q} = 4 \Delta^2 \hat n_{\zeta,\varphi}$ given a probe state of the form~\eqref{general}. The operator $\hat n_{\zeta,\varphi}$ may represent a phase shift or, in the CV limit, a translation, and can involve different combinations of modes $i,j=1,2,3$ depending on the relative weights $\chi_{i,j}$. These weights define collective variables—or, equivalently in the CV limit, effective modes—parametrized by the angles $\zeta$ and $\varphi$. Consequently, for an arbitrary probe state in the SSRC representation—whether single- or multimode, entangled or separable, continuous-variable or fixed-particle-number—we seek the optimal choice of $\chi_{i,j}$ (or equivalently, $\zeta$ and $\varphi$) that maximizes the QFI.

Previous studies have examined the dependence of the QFI on the modal structure in the CV limit~\cite{Gessner:23}, and identified optimal two-mode interferometric manipulations for symmetric states~\cite{PhysRevA.82.012337}. Here, we introduce a general optimization strategy for multimode probes and bosonic systems in arbitrary regimes based on the geometric properties of collective variables, generalizing the findings of~\cite{PhysRevLett.131.030801, PhysRevA.111.052428}. For this, we notice that $\hat n_{\zeta,\varphi}$ can be geometrically seen as a rotation of, say, $\hat n_1$. We can then apply the same rotation to $\hat n_2$ and $\hat n_3$, leading to $\hat n_{\zeta_{2},\varphi_{2}}$ and $\hat n_{\zeta_{3},\varphi_{3}}$, respectively. We have then
\be\label{geo}
\Delta^2 \hat n_{\zeta,\varphi}+\Delta^2\hat n_{\zeta_{2},\varphi_{2}}+\Delta^2\hat n_{\zeta_{3},\varphi_{3}}= \sum_{i=1}^3 \Delta^2 \hat n_i.
\ee
The best strategy to enhance metrological performance when measuring a collective observable $\Delta^2 \hat n_{\zeta,\varphi}$ consists in minimizing the complementary quantity $\Delta^2\hat n_{\zeta_{2},\varphi_{2}}+\Delta^2\hat n_{\zeta_{3},\varphi_{3}}$. The optimal condition $\Delta^2\hat n_{\zeta_{2},\varphi_{2}}+\Delta^2\hat n_{\zeta_{3},\varphi_{3}} = 0$ identifies conserved observables—$\hat n_{\zeta_{2},\varphi_{2}}$ and $\hat n_{\zeta_{3},\varphi_{3}}$—corresponding to underlying physical symmetries~\cite{PhysRevLett.133.260402}.  To illustrate this approach, we consider phase estimation ($\vec n = \vec z$) with a two-mode maximally correlated state, $\ket{{\cal C}_2}=\sum_{\substack{n=0 \\ 2n \leq N}}^N c_n \ket{n}_1 \ket{n}_2 \ket{N - 2n}_3$.
As shown in the Supplemental Material~\cite{SM}, choosing $\cos{\zeta}=\sqrt{2/3}$ and $\sin{\varphi}=\cos{\varphi}=1/\sqrt{2}$ yields $\Delta^2\hat n_{\zeta_{2},\varphi_{2}}+\Delta^2\hat n_{\zeta_{3},\varphi_{3}}=0$ and $\Delta^2 \hat n_{\zeta,\varphi}=9\Delta^2\hat n$, where $\Delta^2\hat n$ is the (identical) photon-number variance of each mode. This result generalizes to a $(k{+}1)$-mode maximally correlated state, $\ket{{\cal C}_{k}}=\sum_{\substack{n=0 \\ 2n \leq N}}^N c_n \ket{n}_1 \ket{n}_2 \dots \ket{N - kn}_{k+1}$, for which ${\cal Q}=4 (k{+}1)^2\Delta^2 \hat n$. Remarkably, the QFI thus scales quadratically with the number of modes, independently of the photon-number variance (that depends on the coefficients $c_n$). 

In the fixed-particle-number regime, where $N$ photons are distributed among $k$ modes, the NOON state $\ket{{\cal C}_k^N}=\frac{1}{\sqrt{2}}(\ket{N}_1\ket{0}_2+\ket{0}_1\ket{N}_2)$ corresponds to the case $k=1$ with $\Delta^2\hat n = N^2/4$, yielding ${\cal Q}=4N^2$. The multimode single-photon regime—where each mode contains one photon—can also be captured by the same general expression by setting $k{+}1=2N$ and $|c_0|=|c_1|=1/\sqrt{2}$: $ \ket{{\cal C}_k^G}=\tfrac{1}{\sqrt{2}}\!\left(\!\prod_{i=1}^{(k{+}1)/2}\!\ket{1}_i+\!\!\prod_{i=(k{+}1)/2+1}^{k{+}1}\!\ket{1}_i\!\right)$, for which $\Delta^2\hat n = 1/4$ and ${\cal Q}=4N^2$.  Although $\ket{{\cal C}_k^N}$ and $\ket{{\cal C}_k^G}$ share the same scaling, their physical origins differ. In $\ket{{\cal C}_k^N}$, only the total photon number is conserved and $\hat n_{\zeta,\varphi}=\hat J_z^{(1,2)}$, so the enhancement arises from single-mode photon-number fluctuations. In contrast, for $\ket{{\cal C}_k^G}$, where $\sum_{i=2}^{k+1} \Delta^2 \hat n_{\zeta_i,\varphi_i}=0$, the relevant operator is $\hat n_{\zeta, \varphi} = \frac{1}{k+1}\sum_{i=1}^{(k+1)/2}(\hat n_i - \hat n_{(k+1)/2+i})$, and the quadratic scaling stems from modal correlations. Despite their distinct physical mechanisms, both states yield identical precision scaling, clarifying the formal equivalence often drawn between them. Additional examples involving single photons with more complex mode structures are discussed in the Supplemental Material.

We now discuss the optimization strategy in the CV limit, where the average photon number—excluding the phase reference, as we have shown to be characteristic of this regime—is $\langle {\cal C}_k|\sum_{i=1}^{k}\hat n_i\ket{{\cal C}_k}=k\overline n$, which defines the shot-noise limit. 

As discussed, evolutions generated by operators such as $\hat J_y^{(i,k+1)}$—that is, involving the phase reference mode—correspond to phase-space translations in the CV limit. We can therefore extend Eq.~(\ref{H}) to the multimode case for $\vec{n}=\vec{y}$, {\it i.e.}, when only $\chi_{i,k+1}\neq 0$ with $i\in\{1,\dots,k\}$. Using the three-mode configuration as an illustrative example (generalized in~\cite{SM}), we have
\begin{eqnarray}\label{P}
&&\hat H = \sum_{i=1}^{2} \chi_{i,3}\hat J_y^{(i,3)} = 
\chi \sum_{i=1}^2 \frac{\chi_{i,3}}{2i\chi} \left( \hat a_i^{\dagger}\hat a_3  - \hat a_i\hat a_3^{\dagger}\right ) \nonumber\\
&&\hspace{1.4cm}= \frac{\chi}{2i} \left ( \hat a_{\zeta,\varphi}^{\dagger}\hat a_3-\hat a_{\zeta,\varphi}\hat a_3^{\dagger}\right ),
\end{eqnarray}
where $\hat a_{\zeta,\varphi}= \frac{1}{2\chi} \left ( \chi_{1,3}\hat a_1+\chi_{2,3}\hat a_2\right ) = \cos{\zeta}\hat a_1 +\sin{\zeta}\hat a_2$ represents a mode transformation~\cite{TrepsModes}. 

Analogously to the phase-estimation scenario, we define $\hat J_y^{(\zeta,\varphi)}= \frac{1}{2i} (\hat a_{\zeta,\varphi}^{\dagger}\hat a_3-\hat a_{\zeta,\varphi}\hat a_3^{\dagger})$, and $\hat J_y^{(\zeta_2,\varphi_2)}=\frac{1}{2i} (\hat a_{\zeta_2,\varphi_2}^{\dagger}\hat a_3-\hat a_{\zeta_2,\varphi_2}\hat a_3^{\dagger})$, where $[\hat a_{\zeta,\varphi},\hat a_{\zeta_2,\varphi_2}^{\dagger}]=0$ and $\hat{a}_{\zeta_2,\varphi_2} = \sin{\zeta} \hat a_1 -\cos{\zeta}\hat  a_2$. We then have 
$\Delta^2 \hat J_y^{(\zeta,\varphi)} + \Delta^2 \hat  J_y^{(\zeta_2,\varphi_2)}= \Delta^2 \hat J_y^{(1,3)} + \Delta^2 \hat  J_y^{(2,3)}$, 
and the precision is maximized by minimizing $\Delta^2 \hat  J_y^{(\zeta_2,\varphi_2)}$. In the CV limit, $\Delta^2 \hat J_y^{(\zeta_i, \varphi_i)} \to N\Delta^2 \hat p_{\zeta_i,\varphi_i}$, with $\hat p_{\zeta_i,\varphi_i}=\frac{i}{\sqrt{2}}(\hat a_{\zeta_i,\varphi_i}^{\dagger}-\hat a_{\zeta_i,\varphi_i})$, $i=1,2$. Hence, our optimization method corresponds to identifying the most suitable mode for a given displacement measurement in the CV regime.

We now compare the previous results and scalings with the case of multimode probes composed of $k$ uncorrelated modes. Such states can be obtained by sending mode~$1$ of a pure state with an explicit phase reference, as in Eq.~\eqref{Psi}, through a sequence of $k-1$ balanced beam splitters—an arrangement typical of multimode interferometers. In this configuration, both the dynamics and measurements involve only the first $k$ modes ({\it i.e.}, excluding the phase reference~\cite{PhysRevA.85.011801}). As shown in~\cite{SM}, the variance of any collective observable satisfies $\Delta^2 \hat n_{\zeta,\varphi} \le k\Delta^2\hat n$, where $\Delta^2\hat n$ denotes the local variance of each mode. In this regime, precision enhancement originates solely from the photon-number fluctuations of the input state, not from modal correlations. Nevertheless, the presence of mode correlations associated with particle entanglement would permit violating the previous inequality, leading to precision enhancement irrespectively of the scaling of the local particle number variance. Interestingly, our result shows that even though mode basis changes - as performed by beam-splitters - can introduce mode entanglement~\cite{Aaron, Lopetegui:25}, such correlations do not enable surpassing the shot-noise limit. 



We conclude by addressing the issue of coupling to the environment and measurement strategies, that can be incorporated to the formalism by adding auxiliary modes. Extending Eq.~\eqref{general} to $\ket{\psi}=\sum_{\substack{n_1,n_2=0 \\ n_1+n_2 \leq N}} c_{n_1,n_2}\ket{n_1}_1\ket{n_2}_2\ket{n_3}_3\ket{n_4}_4\ket{N-\sum_i^4 n_i}_5$. Modes~$4$ and~$5$ act as ancillas or environmental degrees of freedom. Tracing them out models coupling to pure or mixed environments~\cite{LuizMetro,PhysRevA.90.042103,Muller_2011,Barreiro2011AnOQ}, while the coupling itself can be treated as an unknown parameter~\cite{PhysRevResearch.6.013034}.

It is possible to design a measurement strategy reaching the QFI with the following protocol: optimal measurement strategies, for which ${\cal F}_{\theta}={\cal Q}$, are associated with a unitary operator $\hat S$ acting on the probe modes $(1,2,3)$ such that $\hat S\ket{\psi}=\pm\ket{\psi}$ and $\{\hat S,\hat\kappa_1\}=0$~\cite{PhysRevA.79.033822,PhysRevA.108.013707}$,$ where $\hat\kappa_1$ denotes the observable that maximizes ${\cal Q}$ according to the procedure introduced above. By preparing auxiliary modes~$4$ and~$5$ in the entangled state $\tfrac{1}{\sqrt{2}}(\ket{0}_4\ket{1}_5+\ket{1}_4\ket{0}_5)$ one implements a generalized interferometer~\cite{PhysRevA.94.022325} realizing the conditional evolution $\frac{(1+\hat S)}{\sqrt{2}}\,\hat{\cal U}\ket{\psi}\ket{0}_4\ket{1}_5+\frac{(1-\hat S)}{\sqrt{2}}\,\hat{\cal U}\ket{\psi}\ket{1}_4\ket{0}_5$ with $\hat{\cal U}=e^{i\hat\kappa_1\theta}$. The probability of detecting, e.g., outcome $\ket{0}_4\ket{1}_5$ is $p(\theta)=\frac{1}{2}(1+\big\langle\psi\big|\hat{\cal U}\hat S\hat{\cal U}^\dagger\big|\psi\big\rangle)$, and we have ${\cal F}_{\theta}=\frac{1}{p(\theta)}\left ( \frac{\partial p(\theta)}{\partial \theta} \right )^2+\frac{1}{1-p(\theta)}\left ( \frac{\partial p(\theta)}{\partial \theta} \right )^2={\cal Q}$ (see~\cite{SM} for details). Such engineered unitaries and generalized interferometric constructions are routinely implemented in trapped-ion platforms employing motional-state control~\cite{Matsos2024UniversalQG,TrappedionsGKP,Ions}, providing a natural testbed for the present framework.

In conclusion, we have introduced a unified framework that encompasses all known regimes of bosonic quantum metrology—covering symmetric atomic and photonic systems, in both single- and multimode configurations, and in both fixed-photon-number and continuous-variable limits. This model reveals how the QFI fundamentally depends on mode structure and particle statistics, while naturally incorporating noisy dynamics, non-unitary evolutions, and optimal measurement strategies readily implementable in trapped-ion platforms. A key unifying element is the SSRC representation, which enforces total particle-number conservation and clarifies the role of particle entanglement as the essential resource for quantum-enhanced precision. Moreover, our approach provides a general method to optimize precision for arbitrary probes, identifying the conditions under which observables associated with collective operators—analogous to those in Refs.~\cite{PhysRevLett.131.030801, Gross_2012, PhysRevLett.86.4431, PhysRevA.111.052428, Toth2, sorensenManyparticleEntanglementBose2001}—achieve optimal metrological sensitivity in mode-entangled states.

\section*{Acknowledgements}

We acknowledge funding from the Plan France 2030 through the project ANR-22-PETQ-0006 and from ANR-24-CE97-0003 EQUIPPS. We are also indebted to N. Fabre, N. Moulonguet and J. D. Diaz Martinez for inspiring discussions. 

\section{End Matter}
\subsection*{General expression for the variance}
The general expression of the variance $\Delta^2\hat J_{\vec{n}}$, $\vec{n}=\cos{\beta}\vec{z}+\sin{\beta}\cos{\varphi}\vec{x}+\sin{\beta}\sin{\varphi}\vec{y}$ for an arbitrary pure SSRC state (as \eqref{Psi}) is given by
\begin{eqnarray}\label{eqvariance}
&&\Delta^2 \hat J_{\vec{n}} =\frac{N}{4}\sin^2\beta +\frac{1}{2}\left(N\overline{n}-\overline{n^2}\right)\sin^2\beta + \Delta^2 n \cos^2\beta - \nonumber  \\
&&\left(\sum_{n=0}^{N-1}\text{Re}\{c_{n}c_{n+1}^{\ast}e^{-i\varphi}\}\sqrt{\left(n+1\right)\left(N-n\right)}\right)^{2}\sin^2\beta+ \nonumber \\
 && \frac{1}{2}\sum_{n=0}^{N-2}\text{Re}\{c_{n}c_{n+2}^{\ast}e^{-2i\varphi}\} \times \nonumber \\
 &&\sqrt{\left(N-n\right)\left(N-n-1\right)\left(n+1\right)\left(n+2\right)} \sin^2\beta \\
  &&+ \sum_{n=0}^{N-1}\text{Re}\{c_{n}c_{n+1}^{\ast}e^{-i\varphi}\}\left(2n-2\overline{n}+1\right)\times \nonumber \\
  &&\sqrt{\left(N-n\right)\left(n+1\right)}\cos\beta\sin \beta ,\nonumber
\end{eqnarray}
where $\Delta^2 n  = \overline{n^2} - \overline{n}^2$, with $\overline{n^2} = \sum_{n=0}^N |c_n|^2 n^2$ and $\overline{n} = \sum_{n=0}^N |c_n|^2 n$.

\subsection{The CV limit of rotations around the $y$ axis (phase-space translations)}

Rotations around the axis $\vec{y}$ correspond to setting $\beta=\pi/2$ and $\varphi=\pi/2$, so defining $\theta=q/\sqrt{N}$ as in the main text, we have that 
\be \label{x}
\delta \theta=\frac{\delta q}{\sqrt{N}}\geq\frac{1}{\sqrt{4\nu\Delta^2 \hat J_y}},
\ee
where, using \eqref{eqvariance}
\begin{eqnarray}\label{limitx}
&&\Delta^2 \hat J_{y} =\frac{N}{4} +\frac{1}{2}\left(N\overline{n}-\overline{n^2}\right) - \nonumber  \\
&&\left(\sum_{n=0}^{N-1}\text{Re}\{-ic_{n}c_{n+1}^{\ast}\}\sqrt{\left(n+1\right)\left(N-n\right)}\right)^{2}+ \nonumber \\
 && \frac{-1}{2}\sum_{n=0}^{N-2}\text{Re}\{c_{n}c_{n+2}^{\ast}\} \times \nonumber \\
 &&\sqrt{\left(N-n\right)\left(N-n-1\right)\left(n+1\right)\left(n+2\right)}.
\end{eqnarray}
When considering $\delta q$ and $\overline n/N \to 0$ (which is the relevant precision in the CV limit), we have that $\overline n/N \to 0$, $\overline{n^2}/N \to 0$ and 
\be\label{limiteps}
\delta q \geq \frac{1}{\sqrt{4 \nu \Delta^2\hat p}},
\ee
with $\hat p = \frac{i}{2}(\hat a_1-\hat a_1^{\dagger})$. We see that the dependency on $N$ of $\Delta^2 \hat J_y$ disappears. At most, one can expect a linear scaling of precision with $\overline n$. Moreover, we see that $ q$ can be associated with the amplitude of a displacement in phase space, generated by $\hat p$ (see \cite{SM}).

\subsection{The CV limit of rotations around the $z$ axis (phase-space rotations)}

In the CV limit, states are restricted to a tangent, but rotations around the $z$ axis ($\sin{\beta}=0$ and $\cos{\beta}=1$) can be of any angle. Eq. \eqref{eqvariance} then becomes:
\be\label{eq:variance_generalz}
\Delta^2 \hat J_{z} = \Delta^2 \hat n.
\ee
The interest of this results is that it does not depend on the rotation direction or on the fact of being in the CV limit or not.

\onecolumngrid
\appendix

\section{Supplemental Material}

\section{From the SSRC representation to coherent states}

In this section we show how to extract a CV coherent state (Glauber coherent state) $\ket{\alpha}_a$ ($\alpha \in \mathbb{C}$) in a mode $a$  from a Fock state  $\ket{N}_b$ in an orthogonal mode $b$. We will see that it requires to consider the $N\rightarrow \infty$.
The two modes beeing orthogonal their respective anhilation and creation operator fulfill $[\hat{a},\hat{b}^\dagger]=0$. We fix $\alpha\in\mathbb C$. We consider an $N$-and-$\alpha$-dependent linear transformation $\hat U$ of the mode given by
\begin{align*}
    \hat U\hat a^\dagger\hat U^\dagger=\sqrt{1-\frac{\abs{\alpha}^2}{N}}\hat a^\dagger-\frac{\alpha^*}{\sqrt{N}}\hat b^\dagger && \hat U\hat b^\dagger\hat U^\dagger =\frac{\alpha}{\sqrt{N}}\hat a^\dagger+\sqrt{1-\frac{\abs{\alpha}^2}{N}}\hat b^\dagger.
\end{align*}
The transformation $\hat U$ is a rotation $\hat R(\theta,\phi)(\alpha,N))=e^{i\phi \hat J_z}e^{i\theta \hat J_y}$ with parameters $\theta(\alpha,N)=2\arccos(\sqrt{1-\frac{\abs{\alpha}^2}{N}})$ and $\phi(\alpha,N)=\operatorname{arg}(\alpha)$. A quick computation allows us to verify that $[\hat U\hat a^\dagger\hat U^\dagger,\hat U\hat b^\dagger\hat U^\dagger]=0$, such that it indeed corresponds to a unitary transformation. We can now consider in the limit of $N\to\infty$ the state $\hat U\ket{N}_b$
\begin{subequations}
    \begin{align}
        \hat U\ket{N}_b&=\frac{(\hat U\hat b^\dagger\hat U^\dagger)^N}{\sqrt{N!}}\ket{\emptyset}\\
        &=\frac{1}{\sqrt{N!}}\left(\frac{\alpha}{\sqrt{N}}\hat a^\dagger+\sqrt{1-\frac{\abs{\alpha}^2}{N}}\hat b^\dagger\right)^N\ket{\emptyset}\\
        &=\frac{1}{\sqrt{N!}}\sum_{k=0}^N \binom{N}{k}\frac{\alpha^k}{\sqrt{N}^k}\sqrt{1-\frac{\abs{\alpha}^2}{N}}^{N-k} (a^\dagger)^k(\hat b^\dagger)^{N-k}\ket{\emptyset}\\
        &=\sum_{k=0}^N \binom{N}{k}\frac{\sqrt{k!}\sqrt{(N-k)!}}{\sqrt{N!}}\frac{\alpha^k}{\sqrt{N}^k}\sqrt{1-\frac{\abs{\alpha}^2}{N}}^{N-k} \ket{k}_a\ket{N-k}_b\\
        &= \sum_{k=0}^N \sqrt{\frac{N(N-1)\cdots(N-k+1)}{k!}}\frac{\alpha^k}{\sqrt{N}^k}\sqrt{1-\frac{\abs{\alpha}^2}{N}}^{N}\sqrt{1-\frac{\abs{\alpha}^2}{N}}^{-k}\ket{k}_a\ket{N-k}_b\\
        &\simeq \sum_{k=0}^N \frac{\sqrt{N}^k}{\sqrt{k!}}\frac{\alpha^k}{\sqrt{N}^k} e^{-\abs{\alpha}^2/2}\ket{k}_a\ket{N-k}_b\\
        &=e^{-\abs{\alpha}^2/2}\sum_{k=0}^N \frac{\alpha^k}{\sqrt{k!}}\ket{k}_a\ket{N-k}_b.
    \end{align}    
\end{subequations}

Where the approximation is considered term by term: $k$ is fixed while $N\to\infty$.
\begin{align}
    N(N-1)\cdots(N-k+1)\simeq N^k &&\sqrt{1-\frac{\abs{\alpha}^2}{N}}^N\simeq e^{-\abs{\alpha}^2/2} &&\sqrt{1-\frac{\abs{\alpha}^2}{N}}^k\simeq 1.
\end{align}

\section{\label{sec:3mode} Parameter estimation for multimode SSRC state}

In the following, we demonstrate how, starting from a multimode SSRC state one can optimize the quantum Fisher information (QFI) associated with generators acting on different modes, but all associated with a same given rotation axis ${\vec{n}}$. For instance, when $ {\vec{n}}=\vec{z}$ we are analyzing phase estimation using multimode rotations, and this can be done in different ways given a general multimode bosonic state, {\it i.e.}, by differently manipulating each mode so as to optimize precision when measuring a parameter that controls the collective evolution, as will be seen later in this section. This analysis highlights the role of mode structure and correlations between them in enhancing precision, a procedure directly related to entanglement detection using collective variables studied in \cite{PhysRevA.111.052428}. 

We start by considering a general three-mode state conserving the total number of particles $N$, of the form:
\begin{align}
\label{eq:state_3mode}
\lvert\psi\rangle=\sum_{\substack{ n_1,n_2 = 0 \\ n_1+n_2 \leq N }}^{N}c_{n_1,n_2}\lvert n_1\rangle_{1}\lvert n_2\rangle_{2}\lvert N-\left(n_1+n_2\right)\rangle_{3}.
\end{align}
This state undergoes pairwise rotations generated by $\hat{J}_{z}^{(i,j)}$, that encode parameters $\theta_{i,j}$:
\begin{align*}
 \hat{U} = \prod_{\substack {i,j = 1 \\ i<j}}^{3} \hat{U}_{i,j}= \prod_{\substack {i,j = 1 \\ i<j}}^{3}e^{i\hat{J}_{z}^{(i,j)}\theta_{i,j}},
\end{align*}
and the transformed state becomes 
\begin{align}
\label{eq:jz_ij}
\lvert \tilde{\psi}\rangle = \hat{U}\lvert \psi \rangle = \prod_{\substack {i,j = 1 \\ i<j}}^{3}e^{i\hat{J}_{z}^{(i,j)}\theta_{i,j}} \lvert \psi \rangle = e^{\frac{i}{2}\left(\hat{n}_{1}\left(\theta_{1,3}+\theta_{1,2}\right)+\hat{n}_{2}\left(\theta_{2,3}-\theta_{1,2}\right)-\hat{n}_{3}\left(\theta_{2,3}+\theta_{1,3}\right)\right)}\lvert \psi \rangle.
\end{align}
To find each bimodal generator that can be combined to optimize the QFI associated with the estimation of a parameter governing a collective evolution, we start by computing the QFI associated with the general evolution \eqref{eq:jz_ij}. This can be done by parametrizing the rotation angles $\theta_{i,j}$ such that the transformation $\hat{U}$ of eq.~\eqref{eq:jz_ij} becomes equivalent to a global phase evolution under an effective generator $\hat \kappa_3$. To do so we define the following operators:
\begin{subequations}
\label{eq:rotations_3d}
\begin{align}
\label{eq:n_3_jz}
\hat \kappa_3&=\cos\phi_2 \hat{n}_{3}+\sin\phi_2\left(\cos\phi_1 \hat{n}_{2}+\sin\phi_1 \hat{n}_{1}\right),\\
\hat \kappa_2&=\sin\phi_2 \hat{n}_{3}-\cos\phi_2\left(\cos\phi_1 \hat{n}_{2}+\sin\phi_1 \hat{n}_{1}\right),\\
\hat \kappa_1&=\sin\phi_1 \hat{n}_{2}-\cos\phi_1 \hat{n}_{1}, 
\end{align}
\end{subequations}
where the parameters $\phi_i$ and $\zeta$ can be defined as  
\begin{subequations}
\begin{align}
\left(\theta_{1,3}+\theta_{1,2}\right)&=\zeta\sin\phi_1\sin\phi_2,\\
\left(\theta_{2,3}-\theta_{1,2}\right)&=\zeta\sin\phi_2\cos\phi_1,\\
\left(\theta_{2,3}+\theta_{1,3}\right)&=-\zeta\cos\phi_2.
\end{align}
\end{subequations}
With these definitions the expression in Eq.~\eqref{eq:jz_ij} becomes 
\begin{align}
\label{eq:unitary_zeta}
 \hat{U} \lvert \psi \rangle=\prod_{\substack {i,j = 1 \\ i<j}}^{3}e^{i\hat{J}_{z}^{(i,j)}\theta_{i,j}} \lvert \psi \rangle = e^{i\frac{\zeta}{2} \hat \kappa_3}\lvert \psi \rangle.
\end{align}
Thus, the effective generator for the collective transformation involving different mode rotations around the same axis is given by $\hat \kappa_3$. To maximize the precision for estimating $\zeta/2$, we therefore want to maximize the variance $\Delta^2\hat \kappa_3$ of this generator, leading to the quantum Cram\'er-Rao bound:
\begin{align*}
\delta \left(\frac{\zeta}{2}\right) \geq  \frac{1}{\sqrt{ \nu \mathcal{Q}}}= \frac{1}{\sqrt{4 \nu \Delta^{2}\hat \kappa_3}},
\end{align*}
where $\mathcal{Q} = 4\Delta^{2}\hat \kappa_3$ is the quantum Fisher information (QFI), and $\nu$  denotes the number of independent measurements. For readability, we will omit $\nu$ in what follows. 

With the definitions in eq.~\eqref{eq:rotations_3d}, we have the following constraint on the total variance
\begin{align}
\label{eq:variance_max}
\Delta^{2}\hat \kappa_3+\Delta^{2}\hat \kappa_2+\Delta^{2}\hat \kappa_1=\Delta^{2}\hat{n}_{1}+\Delta^{2}\hat{n}_{2}+\Delta^{2}\hat{n}_{3}.
\end{align}
This implies that the total variance $\Delta^{2}\hat \kappa_3+\Delta^{2}\hat \kappa_2+\Delta^{2}\hat \kappa_1$ is fixed. This sets an upper bound on the achievable QFI, and shows that the maximum precision attainable for estimating $\zeta$ -- which corresponds to maximizing $\Delta^{2}\hat \kappa_3$ -- is achieved when $\Delta^{2}\hat \kappa_2+\Delta^{2}\hat \kappa_1 = 0$, and is determined by $\Delta^{2}\hat \kappa_3 = \Delta^{2}\hat{n}_{1}+\Delta^{2}\hat{n}_{2}+\Delta^{2}\hat{n}_{3}$. Notice that satisfying these conditions is not always possible, and in this case, one can always minimize the sums of variances. In practice, this means choosing the collective transformation $\hat \kappa_3$ such that the combined three-mode state has largest collective variance fixed by $\Delta^{2}\hat{n}_{1}+\Delta^{2}\hat{n}_{2}+\Delta^{2}\hat{n}_{3}$. 
We can explicit the variances as follows:
\begin{subequations}
\label{eq:var_3d_z}
\begin{align}
\Delta^{2}\hat \kappa_3=&\cos^{2}\phi_2\Delta^{2}\hat{n}_{3}+\sin^{2}\phi_2\cos^{2}\phi_1\Delta^{2}\hat{n}_{2}+\sin^{2}\phi_2\sin^{2}\phi_1\Delta^{2}\hat{n}_{1}+2\sin^{2}\phi_2\sin\phi_1\cos\phi_1\text{Cov}\{\hat{n}_{1}\hat{n}_{2}\}\\ \nonumber
&+2\sin\phi_2\cos\phi_2\cos\phi_1\text{Cov}\{\hat{n}_{3}\hat{n}_{2}\}+2\sin\phi_2\cos\phi_2\sin\phi_1\text{Cov}\{\hat{n}_{3}\hat{n}_{1}\},\\ 
\Delta^{2}\hat \kappa_2=&\sin^{2}\phi_2\Delta^{2}\hat{n}_{3}+\cos^{2}\phi_2\cos^{2}\phi_1\Delta^{2}\hat{n}_{2}+\cos^{2}\phi_2\sin^{2}\phi_1\Delta^{2}\hat{n}_{1}+2\cos^{2}\phi_2\sin\phi_1\cos\phi_1\text{Cov}\{\hat{n}_{1}\hat{n}_{2}\}\\ \nonumber
&-2\sin\phi_2\cos\phi_2\cos\phi_1\text{Cov}\{\hat{n}_{3}\hat{n}_{2}\}-2\sin\phi_2\cos\phi_2\sin\phi_1\text{Cov}\{\hat{n}_{3}\hat{n}_{1}\},\\
\Delta^{2}\hat \kappa_1=&\sin^{2}\phi_1\Delta^{2}\hat{n}_{2}+\cos^{2}\phi_1\Delta^{2}\hat{n}_{1}-2\sin\phi_1\cos\phi_1\text{Cov}\{\hat{n}_{1}\hat{n}_{2}\},
\end{align}
\end{subequations}
with 
\begin{align*}
&\Delta^{2}\hat{n}_{i} = \overline{n_{i}^{2}}-\overline{n}_{i}^{2}; \quad
\overline{n_{i}^{2}}=\sum_{\substack{ n_1,n_2 = 0 \\ n_1+n_2 \leq N }}^{N}|c_{n_{1},n_{2}}|^{2}n_{i}^{2}; \quad
\overline{n}_{i}=\sum_{\substack{ n_1,n_2 = 0 \\ n_1+n_2 \leq N }}^{N}|c_{n_{1},n_{2}}|^{2}n_{i};\\
&\text{Cov}\{\hat{n}_{i},\hat{n}_{j}\}=\sum_{\substack{ n_1,n_2 = 0 \\ n_1+n_2 \leq N }}^{N}|c_{n_{1},n_{2}}|^{2}n_{i}n_{j}-\overline{n}_{i}\overline{n}_{j}.
\end{align*}
Additionally, using the fact that the total photon number is conserved, {\it i.e.}, $n_3 = N-(n_1+n_2)$, we obtain the following relations:
\begin{align*}
\Delta^{2}\hat{n}_{3} &= \Delta^{2}\hat{n}_{1}+\Delta^{2}\hat{n}_{2}+2\text{Cov}\{\hat{n}_{1}\hat{n}_{2}\},\\
\text{Cov}\{\hat{n}_{3}\hat{n}_{2}\} &= -\Delta^{2}\hat{n}_{2}-\text{Cov}\{\hat{n}_{1}\hat{n}_{2}\},\\
\text{Cov}\{\hat{n}_{3}\hat{n}_{1}\} & = -\Delta^{2}\hat{n}_{1}-\text{Cov}\{\hat{n}_{1}\hat{n}_{2}\}.
\end{align*}
With these, the expressions in eq.~\eqref{eq:var_3d_z} become:
\begin{subequations}
\label{eq:variance_3d_ref}
\begin{align}
\Delta^{2}\hat \kappa_3=& \left(\cos\phi_2-\sin\phi_2\sin\phi_1\right)^{2}\Delta^{2}\hat{n}_{1}+\left(\cos\phi_2-\sin\phi_2\cos\phi_1\right)^{2}\Delta^{2}\hat{n}_{2}\\ \nonumber
&+2\text{Cov}\{\hat{n}_{1}\hat{n}_{2}\}\left(\cos^{2}\phi_2+\sin^{2}\phi_2\sin\phi_1\cos\phi_1-\sin\phi_2\cos\phi_2\left(\cos\phi_1+\sin\phi_1\right)\right)\\
\Delta^{2}\hat \kappa_2=&\left(\cos\phi_2\cos\phi_1+\sin\phi_2\right)^{2}\Delta^{2}\hat{n}_{2}+\left(\sin\phi_2+\cos\phi_2\sin\phi_1\right)^{2}\Delta^{2}\hat{n}_{1}\\ \nonumber
&+2\text{Cov}\{\hat{n}_{1}\hat{n}_{2}\}\left(\sin^{2}\phi_2+\cos^{2}\phi_2\sin\phi_1\cos\phi_1+\sin\phi_2\cos\phi_2\left(\sin\phi_1+\cos\phi_1\right)\right)\\
\Delta^{2}\hat \kappa_1=&\sin^{2}\phi_1\Delta^{2}\hat{n}_{2}+\cos^{2}\phi_1\Delta^{2}\hat{n}_{1}-2\sin\phi_1\cos\phi_1\text{Cov}\{\hat{n}_{1}\hat{n}_{2}\}.
\end{align}
\end{subequations}
Using Eq.~\eqref{eq:variance_max}, we can compute the total variance:
\begin{align}
\label{eq:variance_total_ssr}
\Delta^{2}\hat \kappa_3+\Delta^{2}\hat \kappa_2+\Delta^{2}\hat \kappa_1=2\Delta^{2}\hat{n}_{1}+2\Delta^{2}\hat{n}_{2}+2\text{Cov}\{\hat{n}_{1}\hat{n}_{2}\}.
\end{align}
The precision bound for estimating $\zeta$ becomes
\begin{align}
\label{eq:qfi_3d_gen}
\delta \left(\frac{\zeta}{2}\right)\geq \frac{1}{\sqrt{4\Delta^{2}\hat \kappa_3}} = \frac{1}{\sqrt{4\left( 2\Delta^{2}\hat{n}_{1}+2\Delta^{2}\hat{n}_{2}+2\text{Cov}\{\hat{n}_{1}\hat{n}_{2}\}\right)}} .
\end{align}
This expression shows that the precision depends not only on the photon number variance in each mode but also on their mutual correlations, as expected in the multimode configuration. Nevertheless, such correlations can be casted as the measurement of a single collective mode defined in terms of modes $1$, $2$ and $3$ (Eq. \eqref{eq:n_3_jz}).

We now study as an illustrative example the particular state in which $n_1 = n_2 = n$, {\it i.e.}, the first two modes are maximally correlated, containing the same number of photons, while the reference mode is occupied by $N-2n$ photons, {\it i.e.}, a state in the general form $\ket{\psi}= \sum_{n=0}^{N/2} c_n \ket{n}_1\ket{n}_2\ket{N-2n}_3$. In this case, expression~\eqref{eq:variance_total_ssr} simplifies to
\begin{align}
\Delta^{2}\hat \kappa_3+\Delta^{2}\hat \kappa_2+\Delta^{2}\hat \kappa_1=6\Delta^{2}\hat{n},
\end{align}
where $\hat n=\frac{1}{2}(\hat n_1+\hat n_2)$. Thus, we can maximize $\Delta^{2}\hat \kappa_3$ by minimizing $\Delta^{2}\hat \kappa_2+\Delta^{2}\hat \kappa_1$, with the maximum value being fixed by $6\Delta^{2}\hat{n}$.
For such a state, the variances defined in eq.~\eqref{eq:variance_3d_ref} read as follows: 
\begin{subequations}
\begin{align}
\Delta^{2}\hat \kappa_3=&\left(2\cos\phi_2-\sin\phi_2\left(\sin\phi_1+\cos\phi_2\right)\right)^{2}\Delta^{2}\hat{n}\\
\Delta^{2}\hat \kappa_2=& \left(2\sin\phi_2+\cos\phi_2\left(\sin\phi_1+\cos\phi_1\right)\right)^{2}\Delta^{2}\hat{n}\\
\Delta^{2}\hat \kappa_1=&\left(\sin\phi_1-\cos\phi_1\right)^{2}\Delta^{2}\hat{n}.
\end{align}
\end{subequations}
In order to minimize $\Delta^{2}\hat \kappa_2+\Delta^{2}\hat \kappa_1$, we first choose $\sin\phi_1 = \cos \phi_1 = 1/\sqrt{2}$, which gives:
\begin{align*}
\Delta^{2}\hat \kappa_3=&\left(\sqrt{2}\cos\phi_2-\sin\phi_2\right)^{2}2\Delta^{2}\hat{n}\\
\Delta^{2}\hat \kappa_2=& \left(\sqrt{2}\sin\phi_2+\cos\phi_2\right)^{2}2\Delta^{2}\hat{n}\\
\Delta^{2}\hat \kappa_1=& \, 0.
\end{align*}
Next, by choosing the parameters such that
\begin{align*}
\sqrt{2}\sin\phi_2 = -\cos\phi_2,\\
\sin\phi_2  = \frac{1}{\sqrt{3}}
\end{align*}
we finally obtain:
\begin{align*}
\Delta^{2}\hat \kappa_3=&\left(\sqrt{2}+\frac{1}{\sqrt{2}}\right)^{2}\cos^{2}\phi_2 2\Delta^{2}\hat{n}=9\cos^{2}\phi_2\Delta^{2}\hat{n}=6\Delta^{2}\hat{n}\\
\Delta^{2}\hat \kappa_2=& 0\\
\Delta^{2}\hat \kappa_1=& 0.
\end{align*}
Finally, for the precision of estimating $\zeta$, we have 
\begin{align*}
\delta \left(\frac{\zeta}{2}\right)&\geq\frac{1}{\sqrt{4\Delta^{2}\hat \kappa_3}}=\frac{1}{\sqrt{4\cdot 6\Delta^{2}\hat{n}}}=\frac{1}{\sqrt{4\cdot9\cos^{2}\phi_2\Delta^{2}\hat{n}}}\\
\delta{\zeta}&\geq \frac{1}{3\sqrt{\cos^{2}\phi_2\Delta^{2}\hat{n}}}.
\end{align*}
Alternatively, we can estimate the parameter $\zeta\cos\phi_2= \left(\theta_{2,3}+\theta_{1,3}\right) = \zeta \sqrt{\frac{2}{3}}$, for which the corresponding precision limit is given by:
\begin{align}
\delta \left(\zeta\cos\phi_2\right)&\geq\frac{1}{3\sqrt{\Delta^{2}\hat{n}}}.
\end{align}
This shows that, in practice, even though the phase reference mode is not directly measured and only the local variance is measured, the measured parameter $\theta_{1,3}+\theta_{2,3}$ is relative to the phase reference mode, making its presence essential.

\subsection{Generalization to a $k$-mode state}

We can generalize the above analysis to a  $(k+1)$-mode state, where the $(k+1)th$ mode serves as a phase reference for the remaining $k$ modes. In this case, the state in eq.~\eqref{eq:state_3mode} generalizes to
\begin{align}
\lvert\psi\rangle	=\sum_{\substack{n_{1},\cdots, n_k=0\\\sum_{m=1}^{k}n_{m}\leq N}}^{N}c_{n_{1}\cdots n_{k}}\lvert n_{1}\rangle_{1}\cdots\lvert n_{k}\rangle_{k}\lvert N-\sum_{m}^{k}n_{m}\rangle_{k+1}.
\end{align}
In the $(k+1)$-mode case, the operator $\hat{U}$ of eq.~\eqref{eq:jz_ij} extends to
\begin{align}
\label{eq:jz_ij_k}
\hat{U} = \prod_{\substack{i,j = 1\\ j>i}}^{k+1}e^{i\hat{J}_{z}^{(i,j)}\theta_{i,j}}=e^{i\sum_{i<j}^{k+1}\hat{J}_{z}^{(i,j)}\theta_{i,j}}=e^{\frac{i}{2}\sum_{i=1}^{k+1}\hat{n}_{i}\left(\sum_{j=i+1}^{k+1}\theta_{i,j}-\sum_{j=1}^{i-1}\theta_{j,i}\right)}. 
\end{align}
Similar to the three-mode case discussed earlier, we can define transformations generated by the $\hat{n}_m$ operators involving $k+1$ modes, analogous to the operators defined in eq.~\eqref{eq:rotations_3d}. To do so, we introduce a set of operators $\hat \kappa_{m,\phi}$, which depend on $m$ independent parameters $\phi_m$:
\begin{align}
\hat \kappa_m&=\sin\phi_{m}\hat{n}_{m+1}-\cos\phi_{m}\hat \kappa_{m-1,\phi_{m-1}}^{\perp},
\end{align}
with 
\begin{align}
\hat \kappa_m^{\perp}&=\cos\phi_{m}\hat{n}_{m+1}+\sin\phi_{m}\hat \kappa_{m-1,\phi_{m-1}}^{\perp},
\end{align}
where $m \in \{2,\ldots,k\}$. For $m=1$ and $m=k+1$ we define 
\begin{align}
\hat \kappa_1&=\sin \phi_{1}\hat{n}_{2}-\cos \phi_{1} \hat{n}_{1}\\
{\hat{\vec{n}}}^{\perp}_1&=\sin \phi_{1}\hat{n}_{1}+\cos \phi_{1} \hat{n}_{2},\\
\hat \kappa_{k+1,{\phi_k}}& =\hat \kappa_{k,\phi_k}^{\perp}.
\end{align}
With these definitions, the operators analogous to those in eq.~\eqref{eq:rotations_3d} extend to dimension $(k+1)$ as follows:
\begin{subequations}
\label{eq:vec_n_k}
\begin{align}
\hat \kappa_1&=\sin \phi_{1}\hat{n}_{2}-\cos \phi_{1} \hat{n}_{1}\\
\hat \kappa_m&=\sin\phi_{m}\hat{n}_{m+1}-\cos\phi_{m}\hat \kappa_{m-1,\phi_{m-1}}^{\perp}, \quad \text{for} \; m \in \{2,\cdots,k\}\\
{\hat{\hat {\vec{n}}}}_{k+1,{\phi_k}}& =\hat \kappa_{k,\phi_k}^{\perp}.
\end{align}
\end{subequations}
By expanding these operators in terms of  the number operators $\hat{n}_m$, we obtain
\begin{align}
\nonumber
\hat \kappa_1&=\sin \phi_{1}\hat{n}_{2}-\cos \phi_{1} \hat{n}_{1}\\ \nonumber
\hat \kappa_2&=\sin\phi_{2}\hat{n}_{3}-\cos\phi_{2}\left(\cos\phi_{1}\hat{n}_{2}+\sin\phi_{1}\hat{n}_{1}\right)=\sin\phi_{2}\hat{n}_{3}-\cos\phi_{2}\cos\phi_{1}\hat{n}_{2}-\cos\phi_{2}\sin\phi_{1}\hat{n}_{1}\\ \nonumber
\hat \kappa_{m,\phi_{m}} & =\sin\phi_{m}\hat{n}_{m+1}-\cos\phi_{m}\left(\cos\phi_{m-1}\hat{n}_{m}+\sum_{l=2}^{m-1}\left(\prod_{p=l}^{m-1}\sin\phi_{p}\right)\cos\phi_{l-1}\hat{n}_{l}+\left(\prod_{p=1}^{m-1}\sin\phi_{p}\right)\hat{n}_{1}\right),\; \text{for}\;m\in\{3,\cdots,k\}\\ \label{eq:op_vs_num}
\hat \kappa_{k+1,\phi_{k}} & =\cos\phi_{k}\hat{n}_{k+1}+\sin\phi_{k}\left(\cos\phi_{k-1}\hat{n}_{k}+\sum_{l=2}^{k-1}\left(\prod_{p=l}^{k-1}\sin\phi_{p}\right)\cos\phi_{l-1}\hat{n}_{l}+\left(\prod_{p=1}^{k-1}\sin\phi_{p}\right)\hat{n}_{1}\right)\\ \nonumber
&=\cos\phi_{k}\hat{n}_{k+1}+\sum_{l=2}^{k}\left(\prod_{p=l}^{k}\sin\phi_{p}\right)\cos\phi_{l-1}\hat{n}_{l}+\left(\prod_{p=1}^{k}\sin\phi_{p}\right)\hat{n}_{1}
\end{align}
By estimating the variances of these operators, we can show that the following relation indeed holds:
\begin{align}
\label{eq:variance_total_k}
\sum_{m=1}^{k}\Delta^{2}\hat \kappa_{m,\phi_{m}}+\Delta^{2}\hat \kappa_{k+1,\phi_{k}} = \sum_{m=1}^{k+1}\Delta^{2}\hat{n}_{m}.
\end{align}
By defining the parameters $\zeta$ and $\phi_i$ such that, 
\begin{align}
\label{eq:J_z_reduced}
\zeta \hat \kappa_{k+1,{\phi_k}} = \sum_{i=1}^{k+1}\hat{n}_{i}\left(\sum_{j=i+1}^{k+1}\theta_{i,j}-\sum_{j=1}^{i-1}\theta_{j,i}\right)
\end{align}
the operator in eq.~\eqref{eq:jz_ij_k} becomes
\begin{align}
\prod_{\substack{i,j = 1\\ j>i}}^{k+1}e^{i\hat{J}_{z}^{(i,j)}\theta_{i,j}}=e^{i\frac{\zeta}{2}\hat \kappa_{k+1,{\phi_k}}}.
\end{align}
From the equality in eq.~\eqref{eq:J_z_reduced}, and using the expansion of $\hat \kappa_{k+1,\phi_{k}}$ derived in ~\eqref{eq:op_vs_num}, we can identify the parameters  $\zeta$ and $\phi_k$ as follows:
\begin{align*}
\zeta \cos\phi_{k} &= -\sum_{j=1}^{k}\theta_{j,k+1}\\
\zeta \left(\prod_{p=1}^{k}\sin\phi_{p}\right)\cos\phi_{i-1} &= \sum_{j=i+1}^{k+1}\theta_{i,j}-\sum_{j=1}^{i-1}\theta_{j,i} \, , \quad \text{for} \; i = 2, \ldots, k\\
\prod_{p=1}^{k}\sin\phi_{p} &=  \sum_{j=2}^{k+1}\theta_{1,j}.
\end{align*} 
By maximizing the variance $\Delta^{2}\hat \kappa_{k+1,\phi_{k}}$,  we achieve the highest precision in estimating the parameter $\zeta$.  According to relation~\eqref{eq:variance_total_k}, the maximum of $\Delta^{2}\hat \kappa_{k+1,\phi_{k}}$ is bounded by $\sum_{m=1}^{k+1}\Delta^{2}\hat{n}_{m}$, and this bound is reached when $\sum_{m=1}^{k}\Delta^{2}\hat \kappa_{m,\phi_{m}} = 0$. Additionally, using the fact that $n_{k+1} = N-\sum_{m=1}^k n_m$, we have 
\begin{align}
\nonumber
\Delta^{2}\hat{n}_{k+1}&=\sum_{\substack{n_{1},\cdots, n_k=0\\\sum_{m=1}^{k}n_{m}\leq N}}^{N}|c_{n_{1}\cdots n_{k}}|^{2}\left(N-\sum_{m=1}^{k}{n}_{m}\right)^{2}-\left(\sum_{\substack{n_{1},\cdots, n_k=0\\\sum_{m=1}^{k}n_{m}\leq N}}^{N}|c_{n_{1}\cdots n_{k}}|^{2}\left(N-\sum_{m=1}^{k}{n}_{m}\right)\right)^{2}\\ \nonumber
&=\sum_{\substack{n_{1},\cdots, n_k=0\\\sum_{m=1}^{k}n_{m}\leq N}}^{N}|c_{n_{1}\cdots n_{k}}|^{2}\left(\sum_{m=1}^{k}n_{m}\right)^{2}-\left(\sum_{m=1}^{k}\overline{n}_{m}\right)^2\\ \label{eq:var_k+1}
&=\sum_{m=1}^{k}\Delta^{2}\hat{n}_{m}+2\sum_{m=1}^{k-1}\sum_{p>m}^{k}\text{Cov}\{\hat{n}_{m},\hat{n}_{p}\}=\Delta^{2}\hat{n}_{\rm{col}},
\end{align}
where we define collective photon number operator
\begin{align}
 \hat{n}_{\rm{col}} = \sum_{m=1}^{k}\hat{n}_m,
 \end{align}
 which represents the total number of photons occupying the first $k$ modes. Taking into account this relation, eq.~\eqref{eq:variance_total_k} becomes 
\begin{align*}
\sum_{m=1}^{k}\Delta^{2}\hat \kappa_{m,\phi_{m}}+\Delta^{2}\hat \kappa_{k+1,\phi_{k}} = 2\sum_{m=1}^{k}\Delta^{2}\hat{n}_{m}+2\sum_{m=1}^{k-1}\sum_{p>m}^{k}\text{Cov}\{\hat{n}_{m},\hat{n}_{p}\}=\sum_{m=1}^{k}\Delta^{2}\hat{n}_{m} +\Delta^{2}\hat{n}_{\rm{col}}. 
\end{align*}
Thus, the precision limit for estimating $\zeta$, which is achieved by maximizing  $\Delta^{2}\hat \kappa_{k+1,\phi_{k}}$, is given by 
\begin{align}
\delta \zeta \geq  \frac{1}{\sqrt{2\sum_{m=1}^{k}\Delta^{2}\hat{n}_{m}+2\sum_{m=1}^{k-1}\sum_{p>m}^{k}\text{Cov}\{\hat{n}_{m},\hat{n}_{p}\}}} = \frac{1}{\sqrt{\sum_{m=1}^{k}\Delta^{2}\hat{n}_{m} +\Delta^{2}\hat{n}_{\rm{col}}}}.
\end{align}
This expression generalizes the bound given in eq.~\eqref{eq:qfi_3d_gen} and explicitly shows that the precision limit depends not only on the total number of photons present in the system of interest (i.e., in the first $k$ modes, excluding the phase reference), but also on how those photons are distributed across the $k$ modes.

 
As an example, we consider a situation where $k$ modes are uncorrelated, a situation that can, for instance, correspond to sending a initial state $\ket{\psi}=\sum_{n=0}^N c_n \ket{n}_1\ket{N-n}_2$ into a series of $k-1$ beam-splitters, typical of interferometric set-ups. In this case, we will see that we observe a linear scaling with the number of modes. By changing notation and calling mode $k+1$ the phase reference, we have now a $k+1$ modes state where the first $k$ modes are typically considered as the probe, {\it i.e.}, we can consider evolutions and measurements performed in the $k$ modes only. All the $k$ modes have the same photon number variance $\Delta^2 \hat n$ and the covariance ${\rm Cov}(\hat n_i, \hat n_j)$, $i\neq j$ does not depend on the pairs $i$,$j$ neither. In this case, we can define a collective observable as $\hat \kappa_j = \sum_{i=1}^k u_{i,j}\hat n_i$, where $u_{i,j}$ is an orthogonal matrix. We have then that the variance of any collective observable obeys: $\Delta^2 \hat \kappa_j = \sum_{i, i'=1}^k u_{i,j}u_{i',j}(\langle  \hat n_i\hat n_{i'}\rangle-\langle  \hat n_i\rangle \langle\hat n_{i'}\rangle) =  \Delta^2 \hat n +\sum_{i\neq i'; i, i'=1}^k u_{i,j}u_{i',j}{\rm Cov}(\hat n_1,\hat n_2)\leq k  \Delta^2 \hat n$, where $\Delta^2 \hat n $ is the local variance (in a given mode $i \in \{1,..,k\}$), ${\rm Cov}(\hat n_1,\hat n_2)$ the covariance between any pairs of modes and we have used the Cauchy-Schwarz inequality to establish the final result. Notice that the case of unbalanced beam-splitters can also be contemplated by this model, and we can upper bound $\hat \kappa_j \leq k (\Delta^2 n)_{\rm max}$, where $\Delta^2 n)_{\rm max}$ is the maximal value of the single mode particle number variance.

We now consider a case with highly correlated photons, where each of the first $k$ modes contains the same number of photons: $n_1 = n_2 = \cdots = n_k = n$, and the $(k+1)$th mode contains $n_{k+1} = N-kn$ photons. In this case the covariance between any two distinct modes is
\begin{align*}
\text{Cov}\{\hat{n}_{m},\hat{n}_{p}\}  = \Delta^2 \hat{n}
\end{align*}
for $m,p\leq k$ and the total variance becomes
\begin{align}
\label{eq:var_max_n_eq}
\sum_{m=1}^{k}\Delta^{2}\hat \kappa_{m,\phi_{m}}+\Delta^{2}\hat \kappa_{k+1,\phi_{k}}  &= 2 k \Delta^{2}\hat{n} +2\frac{k(k-1)}{2}\Delta^{2}\hat{n} = k(k+1)\Delta^{2}\hat{n} \\
&= \Delta^{2}\left( k \hat{n}\right)+ k\Delta^{2}\hat{n}  =\Delta^{2}\hat{n}_{\rm{col}} + k\Delta^{2}\hat{n}.
\end{align}
Thus, the precision limit when maximizing $\Delta^{2}\hat \kappa_{k+1,\phi_{k}}$ (i.e., when $\sum_{m=1}^{k}\Delta^{2}\hat \kappa_{m,\phi_{m}} = 0$) becomes 
\begin{align}
 \delta \zeta \geq  \frac{1}{\sqrt{\Delta^{2}\hat{n}_{\rm{col}} + k\Delta^{2}\hat{n}}} =  \frac{1}{\sqrt{ k(k+1)\Delta^{2}\hat{n} }}
\end{align}
Comparing this with the case discussed of independent modes, we see that for such a highly correlated state, the precision scales quadratically with the number of modes. This quadratic scaling arises from estimating the correlations between the modes, given by: $2\sum_{m=1}^{k-1}\sum_{p>m}^{k}\text{Cov}\{\hat{n}_{m},\hat{n}_{p}\}.$

As mentioned earlier, to maximize $\Delta^{2}\hat \kappa_{k+1,\phi_{k}}$,  we impose the condition
\begin{align}
\label{eq:sum_0}
\sum_{m=1}^{k}\Delta^{2}\hat \kappa_{m,\phi_{m}} = 0.
\end{align}  
Below we determine the parameters $\phi_m$ that satisfy this condition. We begin by noting that, for a state with an equal number of photons in each mode, while the operators $\hat{n}_m$ are different, their action on the state is the same for all $m \in \{ 1,\cdots k\}$, thus in the expressions in eq.~\eqref{eq:op_vs_num} we factor these operators by denoting $\hat{n}_1 = \cdots = \hat{n}_k = \hat{n}$ and the expressions simplify to:
\begin{align}
\nonumber
\hat \kappa_1&=\left(\sin \phi_{1}-\cos \phi_{1}\right) \hat{n}\\ \nonumber
\hat \kappa_2&=\left(\sin\phi_{2}-\cos\phi_{2}\left(\cos\phi_{1}+\sin\phi_{1}\right)\right)\hat{n} \\ \nonumber
\hat \kappa_{m,\phi_{m}} & =\left(\sin\phi_{m}-\cos\phi_{m}\left(\cos\phi_{m-1}+\sum_{l=2}^{m-1}\left(\prod_{p=l}^{m-1}\sin\phi_{p}\right)\cos\phi_{l-1}+\prod_{p=1}^{m-1}\sin\phi_{p}\right)\right)\hat{n},\; \text{for}\;m=\{3,\cdots,k-1\}\\ \nonumber
\hat \kappa_{k,\phi_{k}} & =\sin\phi_{k}\left( N-k\hat{n}\right)-\cos\phi_{k}\left(\cos\phi_{k-1}+\sum_{l=2}^{k-1}\left(\prod_{p=l}^{k-1}\sin\phi_{p}\right)\cos\phi_{l-1}+\prod_{p=1}^{k-1}\sin\phi_{p}\right)\hat{n}, \\ \label{eq:n_op_n_eq}
\hat \kappa_{k+1,\phi_{k}} &=\cos\phi_{k}\left( N-k\hat{n}\right)+\left(\sum_{l=2}^{k}\left(\prod_{p=l}^{k}\sin\phi_{p}\right)\cos\phi_{l-1}+\prod_{p=1}^{k}\sin\phi_{p}\right)\hat{n},
\end{align}
where we have also used the fact that $\hat{n}_{k+1} = N-k\hat{n}$. Using these expressions, we find that the condition in eq.~\eqref{eq:sum_0} is satisfied by choosing the parameters $\phi_m$ such that 
\begin{align*}
\sin\phi_{1}&=\cos\phi_{1} = \frac{1}{\sqrt{2}}\\
\sin\phi_{2}& = \cos\phi_{2}\left(\cos\phi_{1}+\sin\phi_{1}\right) = \sqrt{2} \cos\phi_{2} = \sqrt{\frac{2}{3}}\\
\sin\phi_{m}&=\cos\phi_{m}\sqrt{m}=\sqrt{\frac{m}{m+1}}, \quad \text{for} \; m = 3,\ldots, k-1.
\end{align*}
With these choices, all variances $\Delta^2 \hat \kappa_{m,\phi_{m}} $ for  $m = \{1,\ldots, k-1\}$ vanish. For $m=k$ we have: 
\begin{align*}
\hat \kappa_{k,\phi_{k}} &=\sin\phi_{k}\left(N-k\hat{n}\right) -\cos\phi_{k}\sqrt{k}\hat{n} = \sin\phi_{k}N-\left(k\sin\phi_{k}+\cos\phi_{k}\sqrt{k}\right)\hat{n}.
\end{align*}
Requiring the variance $\Delta^2\hat \kappa_{k,\phi_{k}}$ to vanish implies:
\begin{align}
\sin\phi_{k} = -\frac{1}{\sqrt{k}}\cos\phi_{k} = \frac{1}{\sqrt{k+1}}.
\end{align}
With this value of $\phi_k$, the operator $\hat \kappa_{k+1,{\phi_k}}$ becomes:
\begin{align}
\hat \kappa_{k+1,{\phi_k}} =\cos\phi_{k}\left(N-k\hat{n}\right) +\sin\phi_{k}\sqrt{k} \hat{n}=\cos\phi_{k}\left(N-k\hat{n}\right) -\cos\phi_{k} \hat{n}= \cos\phi_{k}\left(N-(k+1)\hat{n}\right),
\end{align}
and the variance is:
\begin{align}
\Delta^2\hat \kappa_{k+1,{\phi_k}} = \cos^2\phi_{k}(k+1)^2\Delta^2\hat{n} = \frac{k}{k+1}(k+1)^2\Delta^2\hat{n} = k(k+1)\Delta^2\hat{n},
\end{align}
which indeed recovers the maximum variance given in~\eqref{eq:var_max_n_eq}.The corresponding precision limit for estimating $\zeta$ is then:
\begin{align}
\zeta&\geq\frac{1}{\sqrt{\left(k+1\right)^{2}\cos^{2}\phi_{k}\Delta^{2}\hat{n}}} = \frac{1}{\sqrt{k(k+1)\Delta^2\hat{n}}}.
\end{align}
If instead we estimate the parameter $\zeta \cos\phi_{k} = -\sum_{j=1}^{k}\theta_{j,k+1}$, the bound becomes: 
\begin{align}\label{quad}
\zeta\cos\phi_k&\geq\frac{1}{\sqrt{\left(k+1\right)^2\Delta^{2}\hat{n}}}.
\end{align}
That is, the QFI for estimating the collective parameter  $\sum_{j=1}^{k}\theta_{j,k+1}$ scales quadratically with the number of modes, including the phase reference mode. This illustrates how entanglement ``replaces" coherence in scaling, and identifies the origin of quadratic scaling for entangled states with fixed total photon number but with a rich entangled modal structure as we will see in the following illustrative example. 

\subsection{Application: correlation between internal degrees of freedom of single photons}

We now use the previous results to analyze the problem of parameter estimation using states with both an external mode structure (as the ket's subscripts in the previous section) and an  internal mode structure, as temporal and frequency modes. For instance, a single photon state that propagates in mode $1$ with a spectral amplitude $f(\omega)$ can be written as $\ket{\psi}=\left (\int f(\omega)\hat a^{\dagger}_1(\omega)d\omega\right )\ket{\emptyset}= \int f(\omega)\ket{\omega}_1d\omega$. Of course, different states can be considered, with many propagation modes (external modes) and frequency modes (internal modes) that can be correlated or separable.  We will study the case of multiple single photon states where each photon occupies a single propagation mode and the frequency modes can be entangled or separable, as in \cite{PhysRevLett.131.030801, MacconeNature}, for instance, or in \cite{lyons_attosecond-resolution_2018, Kwiat, chen_hong-ou-mandel_2019, PhysRevLett.132.193603} on its two-photon version. A general $k$ photon state can thus be written as 
\[
\int d\omega_1...\omega_k f(\omega_1,...,\omega_k)\ket{\omega_1,...,\omega_k}.
\]
In this context, we can model the evolution by replacing  $\theta_{i,j}\hat J_{ {\vec{n}}}^{(i,j)}\to t \int \omega \hat J_{ {\vec{n}}}^{(i,j)}(\omega)d\omega$, where $t$ has dimension of time. It is then possible to perform the same analysis that has been done for $\hat n_i$ in Eqs. \eqref{eq:vec_n_k} with operators $\int \omega \hat n_i(\omega)d\omega=\hat \omega_i$, and the parameter to be estimated is now $t$. If we now inspect the linear scaling of uncorrelated modes and Eq.  \eqref{quad} and consider the estimation of $t$, we must replace $\Delta^2\hat n_i \to \Delta^2\hat \omega_i$. Notice that $\Delta^2\hat \omega_i= \left (\int \prod_{j=1}^k d\omega_j |f(\omega_1,...,\omega_i,...\omega_k)|^2 \omega_i^2\right )-\left (\int \prod_{j=1}^k d\omega_j |f(\omega_1,...,\omega_i,...\omega_k)|^2 \omega_i\right )^2$ in this case, and we can consider for simplicity that  $\Delta^2\hat \omega_i= \Delta^2\hat \omega ~\forall ~i$.  Hence, the case of uncorrelated modes and \eqref{quad} lead to the same scaling found in \cite{PhysRevLett.131.030801} and \cite{MacconeNature} for, respectively, separable states with an internal mode structure and entangled states where internal and external degrees of freedom are maximally correlated, using a different and more general approach.

\section{Parameter Estimation with an Arbitrary Generator on a Multimode State in the CV limit}

\begin{figure}
\includegraphics[scale=0.35]{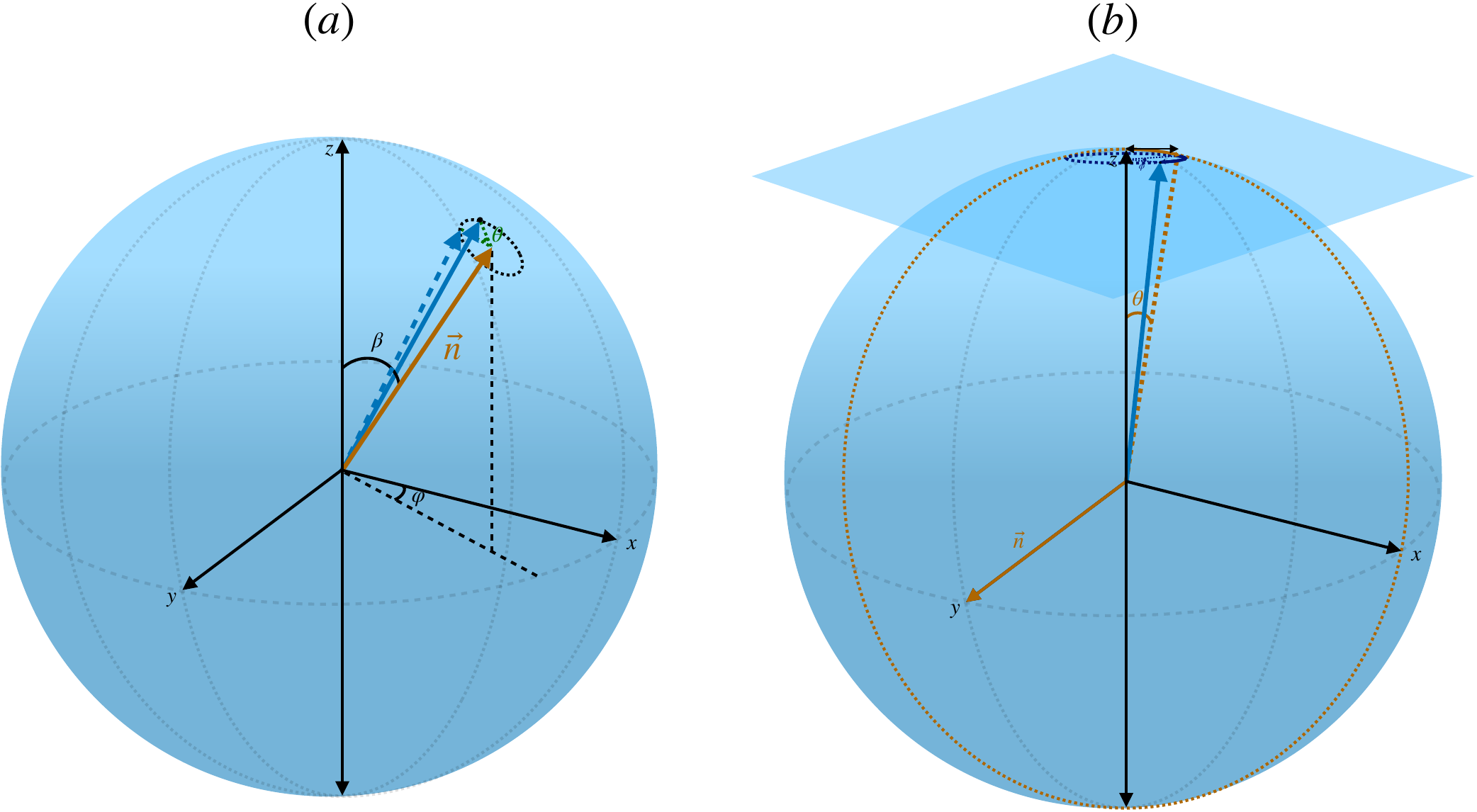}
\caption{ (a) Rotation of the original state (dashed blue line) by an angle $\theta$ around the axis $\hat {\vec{n}}$, defined by polar angle $\beta$ and azimuthal angle $\varphi$. (b) Rotation of the state (blue line) around the $y$-axis by a small angle $\theta$, followed by a rotation around the $z$-axis by an angle $\varphi$. For such a small angle $\theta$, this transformation corresponds to a displacement on the tangent plane. }
\label{fig:fig1}
\end{figure}
We extend the above analysis to a general case where the transformation is performed using a collective operator associated with a rotations around the arbitrary direction ${\vec{n}} = \{ \cos\varphi \sin\beta, \sin\varphi \sin\beta, \cos\beta\}$  with small angles $\theta_{i,j}/\sqrt{N}$, such that $\theta^2_{i,j}/{N}\ll 1$ (see Fig.~\ref{fig:fig1}). For a three-mode system, the total unitary can be written as a product of two-mode rotations:
\begin{align}
\label{eq:op_n_dir}
\prod_{i<j=1}^{3}e^{i\hat{J}_{ {\vec{n}}}^{(i,j)}\frac{\theta_{i,j}}{\sqrt{N}}} = e^{i\hat{J}_{ {\vec{n}}}^{(2,3)}\frac{\theta_{2,3}}{\sqrt{N}}}e^{i\hat{J}_{ {\vec{n}}}^{(1,3)}\frac{\theta_{1,3}}{\sqrt{N}}} e^{i\hat{J}_{ {\vec{n}}}^{(1,2)}\frac{\theta_{1,2}}{\sqrt{N}}}\approx e^{\frac{i}{\sqrt{N}}\left( \hat{J}^{(1,2)}_{ {\vec{n}}}\theta_{1,2}+\hat{J}^{(1,3)}_{ {\vec{n}}}\theta_{1,3}+\hat{J}^{(2,3)}_{ {\vec{n}}}\theta_{2,3}\right) },
\end{align}
where the final expression is obtained taking into account the fact that $\frac{\theta_{i,j}\theta_{i',j'}}{N}\left[ \hat{J}^{(i,j)}_{ {\vec{n}}},\hat{J}^{(i',j')}_{ {\vec{n}}}\right]\ll 1$. We then introduce transformations generated by operators $\hat \kappa_m$ analogous to the ones defined in~\eqref{eq:rotations_3d} with
\begin{subequations}
\label{eq:op_parametrized}
\begin{align}
\hat \kappa_3&=\cos\phi_{2}\hat{J}_{ {\vec{n}}}^{(1,2)}+\sin\phi_{2}\left(\cos\phi_{1}\hat{J}_{ {\vec{n}}}^{(1,3)}+\sin\phi_{1}\hat{J}_{ {\vec{n}}}^{(2,3)}\right),\\
\hat \kappa_2&=\sin\phi_{2}\hat{J}_{ {\vec{n}}}^{(1,2)}-\cos\phi_{2}\left(\cos\phi_{1}\hat{J}_{ {\vec{n}}}^{(1,3)}+\sin\phi_{1}\hat{J}_{ {\vec{n}}}^{(2,3)}\right),\\
\hat \kappa_1&=\sin\phi_{1}\hat{J}_{ {\vec{n}}}^{(1,3)}-\cos\phi_{1}\hat{J}_{ {\vec{n}}}^{(2,3)}.
\end{align}
\end{subequations}
We next define the parameters $\chi$ and $\phi_m$ via the following relations:
\begin{subequations}
\begin{align}
\chi \cos\phi_2 = 2\theta_{1,2},\\
\chi \sin\phi_2\cos\phi_1 = 2\theta_{1,3},\\
\chi \sin\phi_2\sin\phi_1 = 2\theta_{2,3}, 
\end{align}
\end{subequations}
which leads to $\chi^2 = 4\left( \theta^2_{1,2}+\theta^2_{1,3}+\theta^2_{2,3}\right)$. With these definitions, the operator in~\eqref{eq:op_n_dir} can be written as 
\begin{align}
e^{\frac{i}{\sqrt{N}}\left( \hat{J}^{(1,2)}_{ {\vec{n}}}\theta_{1,2}+\hat{J}^{(1,3)}_{ {\vec{n}}}\theta_{1,3}+\hat{J}^{(2,3)}_{ {\vec{n}}}\theta_{2,3}\right) } = e^{i\frac{\chi}{2\sqrt{N}}\hat \kappa_3}.
\end{align}
Thus, the effective generator of the collective transformation is given by $\hat \kappa_3$. By computing the variances of the operators in eq.~\eqref{eq:op_parametrized}, we obtain the following expressions:
\begin{align*}
\Delta^{2}\hat \kappa_3 = & \cos^{2}\phi_{2}\Delta^{2}\hat{J}_{ {\vec{n}}}^{(1,2)}+\sin^{2}\phi_{2}\cos^{2}\phi_{1}\Delta^{2}\hat{J}_{ {\vec{n}}}^{(1,3)}+\sin^{2}\phi_{2}\sin^{2}\phi_{1}\Delta^{2}\hat{J}_{ {\vec{n}}}^{(2,3)}\\
&+\sin\phi_{2}\cos\phi_{2}\Bigg(\cos\phi_{1}\left(\text{Cov}\{\hat{J}_{ {\vec{n}}}^{(1,3)},\hat{J}_{ {\vec{n}}}^{(1,2)}\}+\text{Cov}\{\hat{J}_{ {\vec{n}}}^{(1,2)},\hat{J}_{ {\vec{n}}}^{(1,3)}\}\right)\\
&+\sin\phi_{1}\left(\text{Cov}\{\hat{J}_{ {\vec{n}}}^{(2,3)},\hat{J}_{ {\vec{n}}}^{(1,2)}\}+\text{Cov}\{\hat{J}_{ {\vec{n}}}^{(1,2)},\hat{J}_{ {\vec{n}}}^{(2,3)}\}\right)\Bigg)\\
&+\sin^{2}\phi_{2}\cos\phi_{1}\sin\phi_{1}\left(\text{Cov}\{\hat{J}_{ {\vec{n}}}^{(1,3)},\hat{J}_{ {\vec{n}}}^{(2,3)}\}+\text{Cov}\{\hat{J}_{ {\vec{n}}}^{(2,3)},\hat{J}_{ {\vec{n}}}^{(1,3)}\}\right) \\
\Delta^{2}\hat \kappa_2=&\sin^{2}\phi_{2}\Delta^{2}\hat{J}_{ {\vec{n}}}^{(1,2)}+\cos^{2}\phi_{2}\cos^{2}\phi_{1}\Delta^{2}\hat{J}_{ {\vec{n}}}^{(1,3)}+\cos^{2}\phi_{2}\sin^{2}\phi_{1}\Delta^{2}\hat{J}_{ {\vec{n}}}^{(2,3)}\\
&-\sin\phi_{2}\cos\phi_{2}\Bigg(\cos\phi_{1}\left(\text{Cov}\{\hat{J}_{ {\vec{n}}}^{(1,3)},\hat{J}_{ {\vec{n}}}^{(1,2)}\}+\text{Cov}\{\hat{J}_{ {\vec{n}}}^{(1,2)},\hat{J}_{ {\vec{n}}}^{(1,3)}\}\right)\\
&+\sin\phi_{1}\left(\text{Cov}\{\hat{J}_{ {\vec{n}}}^{(2,3)},\hat{J}_{ {\vec{n}}}^{(1,2)}\}+\text{Cov}\{\hat{J}_{ {\vec{n}}}^{(1,2)},\hat{J}_{ {\vec{n}}}^{(2,3)}\}\right)\Bigg)\\
&+\cos^{2}\phi_{2}\cos\phi_{1}\sin\phi_{1}\left(\text{Cov}\{\hat{J}_{ {\vec{n}}}^{(1,3)},\hat{J}_{ {\vec{n}}}^{(2,3)}\}+\text{Cov}\{\hat{J}_{ {\vec{n}}}^{(2,3)},\hat{J}_{ {\vec{n}}}^{(1,3)}\}\right)\\
\Delta^{2}\hat \kappa_1=&\sin^{2}\phi_{1}\Delta^{2}\hat{J}_{ {\vec{n}}}^{(1,3)}+\cos^{2}\phi_{1}\Delta^{2}\hat{J}_{ {\vec{n}}}^{(2,3)}-\sin\phi_{1}\cos\phi_{1}\left(\text{Cov}\{\hat{J}_{ {\vec{n}}}^{(1,3)},\hat{J}_{ {\vec{n}}}^{(2,3)}\}+\text{Cov}\{\hat{J}_{ {\vec{n}}}^{(2,3)},\hat{J}_{ {\vec{n}}}^{(1,3)}\}\right).
\end{align*}
We can now observe that the sum of the variances of the three collective operators is equal to the sum of the local variances:
\begin{align}
\label{eq:sum_loc_var}
\Delta^{2}\hat \kappa_1 + \Delta^{2}\hat \kappa_2 + \Delta^{2}\hat \kappa_3 = \Delta^{2}\hat{J}_{ {\vec{n}}}^{(1,2)}+ \Delta^{2}\hat{J}_{ {\vec{n}}}^{(1,3)} + \Delta^{2}\hat{J}_{ {\vec{n}}}^{(2,3)}.
\end{align}
Hence, the total variance is conserved and maximizing $\Delta^2\hat \kappa_3$ (and the QFI for $\chi$) is equivalent to minimizing the variances of $\hat \kappa_2$ and $\hat \kappa_2$. 

To estimate these variances, we recall that each $\Delta^2\hat{J}_{ {\vec{n}}}^{(i,j)}$ can be expanded in terms of the $x,y,z$ components of the angular momentum operators:
\begin{align*}
\Delta^{2}\hat{J}_{ {\vec{n}}}^{(i,j)}=& \, \Delta^{2}\hat{J}_{x}^{(i,j)}\cos^{2}\varphi\sin^{2}\beta+\Delta^{2}\hat{J}_{y}^{(i,j)}\sin^{2}\varphi\sin^{2}\beta+\Delta^{2}\hat{J}_{z}^{(i,j)}\cos^{2}\beta\\
&+\left(\text{Cov}\{\hat{J}_{x}^{(i,j)},\hat{J}_{y}^{(i,j)}\}+\text{Cov}\{\hat{J}_{y}^{(i,j)},\hat{J}_{x}^{(i,j)}\}\right)\cos\varphi\sin\varphi\sin^{2}\beta\\
&+\left(\text{Cov}\{\hat{J}_{x}^{(i,j)},\hat{J}_{z}^{(i,j)}\}+\text{Cov}\{\hat{J}_{z}^{(i,j)},\hat{J}_{x}^{(i,j)}\}\right)\cos\varphi\sin\beta\cos\beta\\
&+\left(\text{Cov}\{\hat{J}_{y}^{(i,j)},\hat{J}_{z}^{(i,j)}\}+\text{Cov}\{\hat{J}_{z}^{(i,j)},\hat{J}_{y}^{(i,j)}\}\right)\sin\varphi\sin\beta\cos\beta.
\end{align*}
Summing over all pairs $(i<j)$ gives:
\begin{align*}
\sum_{i<j}^3\Delta^{2}\hat{J}_{ {\vec{n}}}^{(i,j)} =& \, \sum_{i<j}^3\Delta^{2}\hat{J}_{x}^{(i,j)}\cos^{2}\varphi\sin^{2}\beta+ \sum_{i<j}^3\Delta^{2}\hat{J}_{y}^{(i,j)}\sin^{2}\varphi\sin^{2}\beta+ \sum_{i<j}^3\Delta^{2}\hat{J}_{z}^{(i,j)}\cos^{2}\beta\\
&+ \sum_{i<j}^3\left(\text{Cov}\{\hat{J}_{x}^{(i,j)},\hat{J}_{y}^{(i,j)}\}+\text{Cov}\{\hat{J}_{y}^{(i,j)},\hat{J}_{x}^{(i,j)}\}\right)\cos\varphi\sin\varphi\sin^{2}\beta\\
&+ \sum_{i<j}^3\left(\text{Cov}\{\hat{J}_{x}^{(i,j)},\hat{J}_{z}^{(i,j)}\}+\text{Cov}\{\hat{J}_{z}^{(i,j)},\hat{J}_{x}^{(i,j)}\}\right)\cos\varphi\sin\beta\cos\beta\\
&+ \sum_{i<j}^3\left(\text{Cov}\{\hat{J}_{y}^{(i,j)},\hat{J}_{z}^{(i,j)}\}+\text{Cov}\{\hat{J}_{z}^{(i,j)},\hat{J}_{y}^{(i,j)}\}\right)\sin\varphi\sin\beta\cos\beta.
\end{align*}

We can now examine the limit $N\to\infty$, in which mode 3 acts as a classical reference. In this regime, with modes 1 and 2 having finite number of photons, the two-mode coupling operators become continuous-variable quadratures. To demonstrate this, without loss of generality, we consider a particular case where $\hat{J}_{ {\vec{n}}} = \hat{J}_y$, i.e., $\beta = \pi/2$, $\varphi = \pi/2$. 
In this case, the bound for the sum of the variances in~\eqref{eq:sum_loc_var} becomes $\Delta^{2}\hat{J}_{y}^{(1,2)}+ \Delta^{2}\hat{J}_{y}^{(1,3)} + \Delta^{2}\hat{J}_{y}^{(2,3)}$. To compute these variances we first note that for a general $k$-mode state 
\begin{align*}
\lvert\psi\rangle=&\sum_{n_{1}\cdots n_{k}}c_{n_{1},\cdots ,n_{k}}\lvert n_{1}\rangle_{1}\lvert n_{2}\rangle_{2}\cdots\lvert n_{k}\rangle_{k},
\end{align*}
with $\sum_{i=1}^k n_i=N$, the variance of $\hat{J}_y^{(i,j)}$ for any two modes $i,j$ is
\begin{align*}
\Delta^2\hat{J}_y^{(i,j)} =&\, \langle \left(\hat{J}_y^{(i,j)}\right)^2\rangle - \langle \hat{J}_y^{(i,j)} \rangle,
\end{align*}
with
\begin{align*}
\langle\left(\hat{J}_{y}^{(i,j)}\right)^{2}\rangle=&-\frac{1}{4}\langle\psi\rvert\left(\hat{a}_{i}\hat{a}_{j}^{\dagger}\hat{a}_{i}\hat{a}_{j}^{\dagger}-\hat{a}_{i}\hat{a}_{j}^{\dagger}\hat{a}_{i}^{\dagger}\hat{a}_{j}-\hat{a}_{i}^{\dagger}\hat{a}_{j}\hat{a}_{i}\hat{a}_{j}^{\dagger}+\hat{a}_{i}^{\dagger}\hat{a}_{j}\hat{a}_{i}^{\dagger}\hat{a}_{j}\right)\lvert\psi\rangle \\
=& -\frac{1}{4}\sum_{n_{1}\cdots n_{k}}\Bigg(c_{n_{1},\cdots, n_{i}-2,\cdots, n_{j}+2,\cdots, n_{k}}^{\ast}c_{n_{1},\cdots, n_{k}}\sqrt{n_{i}\left(n_{i}-1\right)\left(n_{j}+1\right)\left(n_{j}+2\right)}\\
&+c_{n_{1},\cdots, n_{i}+2,\cdots, n_{j}-2,\cdots, n_{k}}^{\ast}c_{n_{1},\cdots, n_{k}}\sqrt{n_{j}\left(n_{j}-1\right)\left(n_{i}+1\right)\left(n_{i}+2\right)}\Bigg)\\
&+\frac{1}{4}\sum_{n_{1}\cdots n_{k}}|c_{n_{1},\cdots, n_{k}}|^{2}\left(n_{j}\left(n_{i}+1\right)+n_{i}\left(n_{j}+1\right)\right),\\
\langle\left(\hat{J}_{y}^{(i,j)}\right)\rangle=&\,\frac{i}{2}\langle\left(\hat{a}_{i}\hat{a}_{j}^{\dagger}-\hat{a}_{i}^{\dagger}\hat{a}_{j}\right)\rangle\\
=&\,\frac{i}{2}\sum_{n_{1}\cdots n_{k}}\left(c_{n_{1},\cdots, n_{i}-1,\cdots, n_{j}+1,\cdots ,n_{k}}^{\ast}c_{n_{1},\cdots, n_{k}}\sqrt{n_{i}\left(n_{j}+1\right)}-c_{n_{1},\cdots, n_{i}+1,\cdots, n_{j}-1,\cdots, n_{k}}^{\ast}c_{n_{1},\cdots, n_{k}}\sqrt{n_{j}\left(n_{i}+1\right)}\right).
\end{align*}
For the three-mode case,  whenever the rotation involves one of the first two modes and the third (reference) mode, we can analyze the limit $N\to \infty$:
\begin{align*}
\langle\left(\hat{J}_{y}^{(1,3)}\right)^{2}\rangle=& -\frac{1}{4}\sum_{n_{1},n_{2}}\Bigg(c_{n_{1}-2, n_{2}}^{\ast}c_{n_{1},n_{2}}\sqrt{n_{1}\left(n_{1}-1\right)\left(N-(n_{1}+n_2)+1\right)\left(N-(n_{1}+n_2)+2\right)}\\
&+c_{n_{1}+2, n_{2}}^{\ast}c_{n_{1}, n_{2}}\sqrt{\left( N-(n_{1}+n_2)\right)\left(N-(n_{1}+n_2)-1\right)\left(n_{1}+1\right)\left(n_{1}+2\right)}\Bigg)\\
&+\frac{1}{4}\sum_{n_{1}, n_{2}}|c_{n_{1},n_{2}}|^{2}\left(\left( N-(n_{1}+n_2)\right)\left(n_{1}+1\right)+n_{1}\left(N-(n_{1}+n_2)+1\right)\right)\\
\underset{N\to\infty}{\approx}&   -\frac{N}{4}\sum_{n_{1},n_{2}}\Bigg(c_{n_{1}-2, n_{2}}^{\ast}c_{n_{1},n_{2}}\sqrt{n_{1}\left(n_{1}-1\right)}+c_{n_{1}+2, n_{2}}^{\ast}c_{n_{1}, n_{2}}\sqrt{\left(n_{1}+1\right)\left(n_{1}+2\right)}\Bigg)\\
&+\frac{N}{4}\sum_{n_{1}, n_{2}}|c_{n_{1},n_{2}}|^{2}\left(2n_{1}+1\right) =N \langle \left( \frac{i}{2}\left(\hat{a}_1 -\hat{ a}^\dagger_1\right)\right)^2 \rangle = N \langle \, \hat{p}^2_1 \, \rangle \\
\langle\left(\hat{J}_{y}^{(1,3)}\right)\rangle=&\,\frac{i}{2}\sum_{n_{1},n_{2}}\left(c_{n_{1}-1, n_{2}}^{\ast}c_{n_{1}, n_{2}}\sqrt{n_{1}\left(N-(n_1+n_2)+1\right)}-c_{n_{1}+1,n_{2}}^{\ast}c_{n_{1}, n_{2}}\sqrt{\left( N-(n_1+n_2)\right)\left(n_{1}+1\right)}\right)\\
\underset{N\to\infty}{\approx}& \frac{i}{2}\sqrt{N}\sum_{n_{1},n_{2}}\left(c_{n_{1}-1, n_{2}}^{\ast}c_{n_{1}, n_{2}}\sqrt{n_{1}}-c_{n_{1}+1,n_{2}}^{\ast}c_{n_{1}, n_{2}}\sqrt{\left(n_{1}+1\right)}\right) = \sqrt{N} \langle \frac{i}{2}\left(\hat{a}_1 - \hat{a}^\dagger_1\right) \rangle = \sqrt{N} \langle \, \hat{p}_1 \, \rangle ,
\end{align*}
where we have defined the quadrature operator $\hat{p}_1 =\frac{i}{2}\left(\hat{a}_1 - \hat{a}^\dagger_1\right) $, that acts on the approximated state 
\begin{align*}
\lvert\psi\rangle \underset{N\to\infty}{\approx} \sum_{\substack{ n_1,n_2 = 0 \\ n_1+n_2 \leq N }}^{N}c_{n_1,n_2}\lvert n_1\rangle_{1}\lvert n_2\rangle_{2}\lvert N\rangle_{3}.
\end{align*}
Hence:
\begin{align*}
\Delta^2\hat{J}_y^{(1,3)} =N \Delta^2\hat{p}_1.
\end{align*}
Similarly, we can show that 
\begin{align*}
\Delta^2\hat{J}_y^{(2,3)} = N\Delta^2\hat{p}_2,
\end{align*}
with $\hat{p}_2 =\frac{i}{2}\left(\hat{a}_2 - \hat{a}^\dagger_2\right) $. To estimate $\Delta^2\hat{J}_y^{(1,2)}$, we have that 
\begin{align*}
\langle\left(\hat{J}_{y}^{(1,2)}\right)^{2}\rangle =& -\frac{1}{4}\langle\left(\hat{a}_{1}\hat{a}_{2}^{\dagger}-\hat{a}_{1}^{\dagger}\hat{a}_{2}\right)^{2}\rangle\\
=&-\frac{1}{4}\sum_{n_{1},n_2}\Bigg(c_{ n_{1}-2, n_{2}+2}^{\ast}c_{n_{1}, n_{2}}\sqrt{n_{1}\left(n_{1}-1\right)\left(n_{2}+1\right)\left(n_{2}+2\right)}+c_{n_{1}+2,n_{2}-2}^{\ast}c_{n_{1}, n_{2}}\sqrt{n_{2}\left(n_{2}-1\right)\left(n_{1}+1\right)\left(n_{1}+2\right)}\Bigg)\\
&+\frac{1}{4}\sum_{n_{1}, n_{2}}|c_{n_{1}, n_{2}}|^{2}\left(n_{2}\left(n_{1}+1\right)+n_{1}\left(n_{2}+1\right)\right)\\
\langle\left(\hat{J}_{y}^{(1,2)}\right)\rangle=&\,\frac{i}{2}\langle\left(\hat{a}_{1}\hat{a}_{2}^{\dagger}-\hat{a}_{1}^{\dagger}\hat{a}_{2}\right)\rangle=\frac{i}{2}\sum_{n_{1}, n_{2}}\left(c_{n_{1}-1, n_{2}+1}^{\ast}c_{n_{1}, n_{2}}\sqrt{n_{1}\left(n_{2}+1\right)}-c_{ n_{1}+1,n_{2}-1}^{\ast}c_{n_{1}, n_{2}}\sqrt{n_{2}\left(n_{1}+1\right)}\right).
\end{align*}
Using the definitions of $\hat{p}_1, \, \hat{p}_2$ and their conjugates, i.e., 
\begin{align*}
\hat{x}_1 &=\frac{1}{2}\left(\hat{a}_1 + \hat{a}^\dagger_1\right), \\
\hat{x}_2 &=\frac{1}{2}\left(\hat{a}_2 + \hat{a}^\dagger_2\right),
\end{align*}
we obtain
\begin{align*}
\Delta^{2}\hat{J}_{y}^{(1,2)} = \langle\left(\hat{x}_{1}\hat{p}_{2}-\hat{p}_{1}\hat{x}_{2}\right)^{2}\rangle-\left(\langle\hat{x}_{1}\hat{p}_{2}-\hat{p}_{1}\hat{x}_{2}\rangle\right)^{2}. 
\end{align*}
For the total variance we have 
\begin{align*}
\Delta^2\hat{J}_y^{(1,3)}+\Delta^2\hat{J}_y^{(2,3)}+\Delta^2\hat{J}_y^{(1,2)} &= N \Delta^2\hat{p}_1 + N \Delta^2\hat{p}_2 + \langle\left(\hat{x}_{1}\hat{p}_{2}-\hat{p}_{1}\hat{x}_{2}\right)^{2}\rangle-\left(\langle\hat{x}_{1}\hat{p}_{2}-\hat{p}_{1}\hat{x}_{2}\rangle\right)^{2}\\
&= N \left(\Delta^2\hat{p}_1 +  \Delta^2\hat{p}_2 +\frac{1}{N} \left( \langle\left(\hat{x}_{1}\hat{p}_{2}-\hat{p}_{1}\hat{x}_{2}\right)^{2}\rangle-\left(\langle\hat{x}_{1}\hat{p}_{2}-\hat{p}_{1}\hat{x}_{2}\rangle\right)^{2}\right)\right) \\
&\underset{N\to\infty}{\approx} N \left(\Delta^2\hat{p}_1 +  \Delta^2\hat{p}_2 \right).
\end{align*}
Thus, for estimating the variable $\chi$, we obtain:
\begin{align*}
\delta \left(\frac{\chi}{2\sqrt{N}}\right)\geq \frac{1}{\sqrt{4\Delta^2 \hat \kappa_3}} = \frac{1}{2\sqrt{ N \left(\Delta^2\hat{p}_1 +  \Delta^2\hat{p}_2 \right)}}\\
\delta \chi =\delta \left( 2\sqrt{\left( \theta^2_{1,2}+\theta^2_{1,3}+\theta^2_{2,3}\right)}\right)\geq \frac{1}{\sqrt{  \left(\Delta^2\hat{p}_1 +  \Delta^2\hat{p}_2 \right)}}.
\end{align*}
Consequently, in the limit $N\to\infty$, the 3-mode estimation problem reduces to a two-mode continuous-variable problem.  The highly populated third mode serves as a classical reference frame, while the remaining modes are described by canonical quadratures $\hat{x}_i$, $\hat{p}_i$. In this regime, the QFI accumulates contributions from each mode and the estimated parameter not only describes the rotation angle $\theta_{1,2}$ between these two modes but also the angles characterizing their rotation with respect to the reference mode.

At this limit, we can reformulate the collective generators in~\eqref{eq:op_parametrized} using effective quadrature operators. Specifically, in the limit $N \to \infty$ the operators  $\hat{J}_y^{(1,3)}$ and $\hat{J}_y^{(2,3)}$ become (up to a normalization) proportional to the quadratures $\hat{p}_1$ and $\hat{p}_2$ of the first two modes:
\begin{subequations}
\label{eq:rotation_cv}
\begin{align}
\hat \kappa_{2,{\eta}}& = \cos\eta \, \frac{\hat{J}_{y}^{(1,3)}}{\sqrt{N}}+\sin\eta \, \frac{\hat{J}_{y}^{(2,3)}}{\sqrt{N}} \underset{N\to\infty}{\approx}\cos\eta \, \hat{p}_{1}+\sin\eta \, \hat{p}_{2}, \\
\hat \kappa_{1,{\eta}}& = \sin\eta \, \frac{\hat{J}_{y}^{(1,3)}}{\sqrt{N}}-\cos\eta \, \frac{\hat{J}_{y}^{(2,3)}}{\sqrt{N}}  \underset{N\to\infty}{\approx} \sin\eta \, \hat{p}_{1}-\cos\eta  \, \hat{p}_{2}.
\end{align}
\end{subequations}
Then the collective generator in eq.~\eqref{eq:op_n_dir} becomes 
\begin{align}
\label{eq:uni_cv}
\hat{U} = e^{i\frac{\chi}{2}\hat \kappa_{2,{\eta}}},
\end{align}
and the variances satisfy the conserved sum
\begin{align}
\label{eq:var_cv}
\Delta^2 \hat \kappa_{2,{\eta}} + \Delta^2 \hat \kappa_{1,{\eta}} = \Delta^2 \hat{p}_1 + \Delta^2 \hat{p}_2.
\end{align}
Therefore, to maximize the sensitivity (i.e., maximize the QFI associated with $\chi$) we minimize the orthogonal variance $\Delta^2 \hat \kappa_{1,{\eta}} = 0$, since the precision limit associated with Eq. \eqref{eq:uni_cv} becomes.  
\begin{align*}
\delta \left(\frac{\chi}{2}\right)\geq \frac{1}{2\sqrt{\Delta^2 \hat \kappa_{2,{\eta}}}}.
\end{align*}
As an example, we apply the above framework to the case of two modes, with each mode described by its associated quadrature operator $\hat{p}_1$ and $\hat{p}_2$. We define the following collective operators: 
\begin{align}
\hat{p}_+ &= \frac{1}{\sqrt{2}} \left( \hat{p}_1 + \hat{p}_2\right)\\
\hat{p}_- &= \frac{1}{\sqrt{2}} \left( \hat{p}_1 - \hat{p}_2\right),
\end{align}
Then the rotated operators in eq.~\eqref{eq:rotation_cv} become
\begin{align}
\hat \kappa_{2,{\eta}}& = \cos\eta' \hat{p}_{+}+\sin\eta' \, \hat{p}_{-}, \\
\hat \kappa_{1,{\eta}}& = \cos\eta' \, \hat{p}_{-}-\sin\eta'  \, \hat{p}_{+}, 
\end{align}
where $\cos\eta' = \frac{1}{\sqrt{2}}\left(  \sin\eta + \cos\eta\right)$ and $\sin\eta' = \frac{1}{\sqrt{2}}\left(  \cos\eta - \sin\eta\right)$. With this, the unitary~\eqref{eq:uni_cv} is
\begin{align}
\hat{U} = e^{i\frac{\chi}{2}\left( \cos\eta' \hat{p}_{+}+\sin\eta' \, \hat{p}_{-}\right)}, 
\end{align}
and the total variance~\eqref{eq:var_cv} is still conserved:
\begin{align}
\Delta^2 \hat \kappa_{2,{\eta}} + \Delta^2 \hat \kappa_{1,{\eta}}  = \Delta^2 \hat{p}_+ + \Delta^2 \hat{p}_- .
\end{align}
For states for which $ \Delta^2 \hat{p}_-=0$, for instance, the best choice of evolution direction corresponds to  $\sin\eta = \cos\eta =1/\sqrt{2}$,  $ \hat \kappa_{2,{\eta}} =\hat{p}_+ $, and the bound simplifies to: 
\begin{align*}
\delta {\chi} \geq \frac{1}{\sqrt{\Delta^2 \hat{p}_+ }}.
\end{align*}
which is optimized for the given state. This shows that sensitivity to displacements along $\hat p_+$ improves with increasing $\Delta^2 \hat{p}_+$, which, due to the Heisenberg uncertainty principle, can be achieved by squeezing the quadrature  $\frac{1}{\sqrt{2}}(\hat{x}_1+\hat{x}_2)$ of the mode $\hat a_+=\frac{1}{\sqrt{2}}(\hat a_1+\hat a_2)$. This result can be generalized for different number of modes. Hence, in the CV limit, a correct mode choice for the evolution may improve the sensitivity to displacements, as studied in \cite{Gessner:23, TrepsUltimate}

\section{Generalized measurement and optimal measurement strategy}

In this section, following the results of Ref.~\cite{descamps2023timefrequency}, we discuss a scheme allowing to optimally estimate the parameter $\zeta$, discussed in Sec.~\ref{sec:3mode}, for the probe state defined in~\eqref{eq:state_3mode}. For such a 3-mode state, the scheme involves adding two auxiliary modes (mode 4 and 5) that are entangled, such that we have $\lvert \textrm{aux} \rangle = \frac{1}{\sqrt{2}}\left( \lvert 0\rangle_4 \lvert 1\rangle_5 + \lvert 1\rangle_4 \lvert 0\rangle_5\right)$ and the state becomes 
\begin{align*}
\lvert \Psi \rangle = \lvert \psi \rangle \otimes \lvert \textrm{aux} \rangle. 
\end{align*}
To extract information about the parameter $\zeta$ encoded via a unitary $\hat{U}$ defined in~\eqref{eq:unitary_zeta} we apply $\hat{U}$ in a conditional way, i.e.,  $\hat{U}$ is applied to the probe if the auxiliary is in $ \lvert 1\rangle_4 \lvert 0\rangle_5$, and $\hat{S}\hat{U}$ if in $ \lvert 0\rangle_4 \lvert 1\rangle_5 $. Here we defined  a unitary operator $\hat{S}$ such that  $\hat{S} \lvert \psi \rangle = \pm \lvert \psi \rangle$ and it satisfies the condition $\left[\hat{S},\hat{U}\right] \neq 0$. The total evolved state becomes: 
\begin{align}
\label{eq:full_state_aux}
\lvert \tilde{\Psi} \rangle = \frac{1}{\sqrt{2}}\left( \hat{U}\lvert \psi \rangle \otimes \lvert 1\rangle_4 \lvert 0\rangle_5 + \hat{S}\hat{U}\lvert \psi \rangle \otimes \lvert 0\rangle_4 \lvert 1\rangle_5\right)
\end{align}
We now apply a beam splitter-like transformation to auxiliary modes, where in one arm of the balanced beam splitter the input is the mode $ \lvert 1\rangle_4 \lvert 0\rangle_5 $ and on the other $ \lvert 0\rangle_4 \lvert 1\rangle_5$, thus after the beam splitter we obtain 
\begin{align*}
 \lvert 1\rangle_4 \lvert 0\rangle_5 \to \frac{1}{\sqrt{2}}\left( \lvert 1\rangle_4 \lvert 0\rangle_5  + \lvert 0\rangle_4 \lvert 1\rangle_5 \right)\\
 \lvert 0\rangle_4 \lvert 1\rangle_5 \to \frac{1}{\sqrt{2}}\left( \lvert 1\rangle_4 \lvert 0\rangle_5  - \lvert 0\rangle_4 \lvert 1\rangle_5 \right),
\end{align*}
and the full state~\eqref{eq:full_state_aux} after the beam splitter becomes 
\begin{align*}
\lvert \tilde{\Psi} \rangle = \frac{1}{{2}}\left( \left(\hat{U}+\hat{S}\hat{U}\right)\lvert \psi \rangle \otimes \lvert 1\rangle_4 \lvert 0\rangle_5 + \left(\hat{U}-\hat{S}\hat{U}\right)\lvert \psi \rangle \otimes \lvert 0\rangle_4 \lvert 1\rangle_5\right).
\end{align*}
Finally, we make a measurement on a basis of the auxiliary modes, then the probability of detecting $\lvert 1\rangle_4 \lvert 0\rangle_5$ is 
\begin{align*}
p_{10}& = \frac{1}{4} \langle \psi \rvert  \left(U^\dagger+U^\dagger S^\dagger \right) \left(U+SU\right)\lvert \psi \rangle = \frac{1}{4} \langle \psi \rvert  \left(2I+U^\dagger SU+ U^\dagger S^\dagger U\right) \lvert \psi \rangle \\
&= \frac{1}{2} \langle \psi \rvert  \left(I+\textrm{Re}\left\{U^\dagger SU\right\}\right) \lvert \psi \rangle =  \frac{1}{2}+\frac{1}{2}\langle \psi \rvert \textrm{Re}\left\{U^\dagger SU\right\}\lvert \psi \rangle,
\end{align*}
where $I$ is the identity operator and $\textrm{Re}\{ \cdot\}$ denotes the real part of the operator. Similarly,  the probability of detecting $\lvert 0\rangle_4 \lvert 1\rangle_5$ is 
\begin{align*}
p_{01}& = \frac{1}{2}-\frac{1}{2}\langle \psi \rvert \textrm{Re}\left\{U^\dagger SU\right\}\lvert \psi \rangle. 
\end{align*}
Recalling that we study a unitary given by $\hat{U} =  e^{i\frac{\zeta}{2} \hat \kappa_3}$,we refer to the results of Ref.~\cite{descamps2023timefrequency}, which show that for a measurement to be the optimal one, the measurement operator $\hat{S}$ must anticommute with the generator of the unitary: $\{\hat \kappa_3, \hat{S} \} =\hat \kappa_3\,\hat{S} +  \hat{S} \, \hat \kappa_3= 0$. This anticommutation condition ensures that  $\hat{S}$ is an optimal observable for estimating the parameter $\zeta$ in the sense that the measurement of $\hat{S}$ on the evolved state extracts the maximum amount of Fisher information, i.e., the quantum Fisher information.

 Expanding this using the expression from Eq.~\eqref{eq:n_3_jz}, we obtain:
\begin{align*}
\{\hat \kappa_3, \hat{S} \} = \cos\phi_2 \{\hat{n}_{3},\hat{S}\}+\sin\phi_2\left(\cos\phi_1 \{\hat{n}_{2},\hat{S}\}+\sin\phi_1 \{\hat{n}_{1},\hat{S}\}\right) = 0.
\end{align*}
Using photon number conservation $\hat{n}_3 = N-\left( \hat{n}_1+\hat{n}_2\right)$, the above expression becomes 
\begin{align}
\label{eq:anticom}
\left( \cos\phi_2 - \sin\phi_2 \cos\phi_1\right) \{\hat{n}_{2},\hat{S}\} + \left( \cos\phi_2 - \sin\phi_2 \sin\phi_1\right) \{\hat{n}_{1},\hat{S}\} = 2N\cos\phi_2 \hat{S}.
\end{align}

If we consider the special case where the first two modes contain equal photon numbers, ${n}_1 = {n}_2 = n$, and use the parameters  $\sin{\phi_1} =\cos{\phi_1} = 1/\sqrt{2}$, $\sin\phi_2 = 1/\sqrt{3}$ and $\cos\phi_2 = -\sqrt{2/3}$, as derived in Sec.~\ref{sec:3mode}, the condition \eqref{eq:anticom} simplifies to 
\begin{align}
\label{eq:anticom_neq}
 \{\hat{n},\hat{S}\}  = \frac{2}{3}N \hat{S}.
\end{align}
By finding the operator $\hat{S}$ satisfying this condition, we ensure that the QFI derived in Sec.~\ref{sec:3mode}, corresponding to $\Delta^2\hat \kappa_3$ is attained. 

As an  example, consider  a symmetry operator $\hat{S}$ that acts on the first two modes: 
\begin{align}
\label{eq:op_S}
\hat{S} = \sum_{n=0}^{\lfloor N/2 \rfloor} s_n \, \lvert n \rangle_1 \lvert n \rangle_2 \langle n \rvert_1 \langle n \rvert_2 ,
\end{align}
where $s_n = (-1)^{f(n)}$, where $f \in {\cal F}(\mathbb{N},\mathbb{N})$, such that the condition $\hat{S} \lvert \psi \rangle = \pm \lvert \psi \rangle$ is satisfied. One example of such a function is $f(n)=n$. Computing the anticommutator:
\begin{align}
 \{\hat{n},\hat{S}\} =  \sum_{n=0}^{\lfloor N/2 \rfloor} s_n \, \big(\hat{n} \lvert n \rangle_1 \lvert n \rangle_2 \langle n \rvert_1 \langle n \rvert_2 + \lvert n \rangle_1 \lvert n \rangle_2 \langle n \rvert_1 \langle n \rvert_2 \hat{n} \big) =  \sum_{n=0}^{\lfloor N/2 \rfloor} 2s_n n \,  \lvert n \rangle_1 \lvert n \rangle_2 \langle n \rvert_1 \langle n \rvert_2,
\end{align}
and comparing it with the right-hand side of eq.~\eqref{eq:anticom_neq}:
\begin{align}
 \frac{2}{3}N \hat{S} =  \frac{2}{3}N  \sum_{n=0}^{\lfloor N/2 \rfloor} s_n \, \lvert n \rangle_1 \lvert n \rangle_2 \langle n \rvert_1 \langle n \rvert_2,
\end{align}
we find that the condition~\eqref{eq:anticom_neq} is satisfied when $n = \frac{N}{3}$. We note that in systems where the reference frame corresponds to the limit $N \to \infty$, this condition on $n$ becomes unphysical.

\bibliography{bibMetroSSRC}

\begin{thebibliography}{89}
\bibitem [{Not()}]{Note}%
  \BibitemOpen
  \href@noop {} {\bibinfo {title} {where we supposed for simplifying reasons
  and without loss of generality, as we will discuss later, that the two modes
  have the same frequency $\omega$}}\BibitemShut {NoStop}%
\end{thebibliography}

\end{document}